\documentclass[journal,twoside,web]{ieeecolor}

\usepackage{generic}
\usepackage{amsmath,amssymb,amsfonts}
\usepackage{algorithmic}
\usepackage{graphicx}
\usepackage{textcomp}
\usepackage{mathrsfs}
\usepackage[dvipsnames]{xcolor}
\usepackage{cases}
\usepackage{mathtools}
\usepackage{caption}
\usepackage{makeidx}
\usepackage{cuted}
\usepackage{float}
\usepackage{hyperref}
\usepackage{dblfloatfix}

\usepackage{amsthm}

\newtheorem{thm}{Theorem}[section]
\newtheorem{lem}[thm]{Lemma}
\newtheorem{prop}[thm]{Proposition}
\newtheorem{cor}{Corollary}
\newtheorem{conj}{Conjecture}[section]
\theoremstyle{definition}
\newtheorem{defn}{Definition}[section]
\newtheorem{exmp}{Example}[section]
\newtheorem{rem}{Remark}
\newtheorem{assumption}{Assumption} 

\newtheorem*{prop*}{Proposition}
\newtheorem*{thm*}{Theorem}

\def\BibTeX{{\rm B\kern-.05em{\sc i\kern-.025em b}\kern-.08em
    T\kern-.1667em\lower.7ex\hbox{E}\kern-.125emX}}
\begin{document}
\title{Optimal estimation in spatially distributed systems: how far to share measurements from?
}
\author{Juncal Arbelaiz, \IEEEmembership{Member, IEEE},  Bassam Bamieh, \IEEEmembership{Fellow, IEEE},   Anette E.  Hosoi 
and Ali Jadbabaie,  \IEEEmembership{Fellow, IEEE}
\thanks{
J. Arbelaiz acknowledges the support of a Hugh Hampton Young Memorial fellowship from the OGE at MIT and from the Schmidt Science Fellowship, in
partnership with the Rhodes Trust.}
\thanks{
J. Arbelaiz was with the department of Applied Mathematics at the Massachusetts Institute of Technology when she developed this work;  she is now at the Center for Statistics and Machine Learning at Princeton University,  NJ,  08542,  USA (e-mail: jarbelaiz@schmidtsciencefellows.org). }
\thanks{B.  Bamieh is with the department of Mechanical Engineering at the University of California at Santa Barbara,  CA,  93106-2014 USA (e-mail: bamieh@ucsb.edu).}
\thanks{Anette E.  Hosoi is with the department of Mechanical Engineering at the Massachusetts Institute of Technology,  Cambridge,  MA, 02139 USA (e-mail: peko@mit.edu).}
\thanks{Ali Jadbabaie is with the department of Civil Engineering at the Massachusetts Institute of Technology, Cambridge, MA, 02139 USA (e-mail: jadbabai@mit.edu).}}

\maketitle

\begin{abstract}
We consider the centralized optimal estimation problem in spatially distributed systems. We use the setting of spatially invariant systems as an idealization for which concrete and detailed results are given. Such estimators are known to have a degree of spatial localization in the sense that the estimator gains decay in space, with the spatial decay rates serving as a proxy for how far measurements need to be shared in an optimal distributed estimator.  In particular, we examine the dependence of spatial decay rates on problem specifications such as system dynamics, measurement and process noise variances, as well as their spatial autocorrelations. We propose non-dimensional parameters that characterize the decay rates as a function of problem specifications. In particular, we find an 
interesting matching condition between the characteristic lengthscale of the dynamics and the measurement noise correlation lengthscale for which the optimal centralized estimator is  completely decentralized. A new graphical technique --- termed the Branch Point Locus --- is introduced to quantify spatial decay rates in terms of analyticity regions in the complex spatial frequency plane. Our results are illustrated through two case studies of systems with dynamics modeled by diffusion  and the Swift-Hohenberg  equation, respectively. 
\end{abstract}



\section{Introduction}
\label{sec:introduction}

The celebrated \textit{Kalman filter} provides the optimal state estimate  --- in the sense of estimation error variance minimization --- of a linear system in a time recursive manner, under certain technical assumptions.  Since the seminal work of Kalman and Bucy  \cite{Kalman_Bucy:1961, Kalman:1960} a vast amount of literature has followed,  offering generalizations of their  result and providing new insights into the estimation problem \cite{Kailath1974}.  Kalman filters have been widely applied in diverse fields for decades. 
Despite this success,  some practical challenges remain regarding its implementation,  one of them being its application to large-scale spatially distributed systems.  Kalman filters are optimal under the implicit assumption that all-to-all (i.e.,  centralized) 
communications 
are available within the system.  This requires a \textit{centralized fusion center} for the measurements \cite{Das2017, Cattivelli2010}.  However,  centralized communications are often not feasible in large-scale systems,  and even if they are,  communication delays can considerably degrade the performance of a centralized filtering architecture.  Furthermore,  these architectures are fragile to central node failure.  Such challenges make centralized estimation inadequate to large-scale applications \cite{Wang} and motivate the active research field of \textit{distributed Kalman filtering}.



The aim of this work is to characterize the spatial localization of Kalman filters for spatially distributed systems. Specifically, how the interplay between system parameters and dynamics, and the strength of the noises perturbing the plant determines the spatial behavior of the filter: \textit{for a fixed spatial site $x$ in the system, how far do measurements need to be collected from and communicated to site $x$ for state estimation at $x$? Does the strength of the noises in the system affect that?}
This characterization sheds light on \textit{(1)} the amenability of plants and different parameter regimes to decentralized filter  implementations, and \textit{(2)} the structure of decentralized filter architectures that are appropriate for the system at hand.
We carry out this analysis for a particular class of systems with spatiotemporal dynamics over unbounded spatial domains. Our problem is an instance of \textit{distributed-parameter} or \textit{infinite-dimensional filtering}  \cite{Curtain:1975, Curtain:1978, Bala:1981}. In particular, we focus on \textit{spatially invariant systems (SIS)} \cite{Bamieh:2002}. 
SIS are characterized by their dynamics remaining invariant to a notion of translation in the spatial coordinate. 
Under some assumptions \cite{Curtain2010},  SIS provide a useful mathematical abstraction to model large scale arrays of identical subsystems,  regularly arranged in space and with distributed control and sensing capabilities. Applications include smart structures (i.e., large arrays of micro-electro mechanical systems), modular robotics,  formations of autonomous vehicles, and industrial flat sheet processes (e.g., plastic or paper sheet manufacturing). The analysis and synthesis problems for systems with finite spatial domains and boundary conditions can sometimes be addressed by studying spatially invariant systems, using embedding techniques and correction terms -- see e.g., \cite{Langbort2006,Epperlein2016} and references therein. This, together with their analytical tractability, make SIS of special interest within the class of spatially distributed systems.  
The work \cite{Bamieh:2002, Curtain:2011} characterized the structural properties of optimal controllers for SIS.  Related research has followed since then: many have focused on conceiving tractable design strategies for optimal distributed control of SIS  e.g.,  \cite{DeCastro2002,Bamieh:2005,Dhingra2016, Arbelaiz2021,DAndrea2003}; others on extensions to more general classes of systems,  including generalizations to multiple dimensions and heterogeneous systems \cite{Rice2011},  time-varying SIS \cite{Djouadi2019},  and systems with spatially decaying operators \cite{ Motee:2008,Curtain2009_comments,Motee2017}.  However,  only a few have analyzed the Kalman filtering problem for particular examples of SIS (e.g., \cite{Henningsson2007,Massioni2011,Arbelaiz:2020,Arbelaiz2022}).
In this work, we study Kalman filters for SIS in more  generality.
We assume that measurements are distributed in the spatial domain,  but corrupted by (spatially and temporally) stationary measurement noise.  In this setting,  we are concerned with the asymptotic behavior of the Kalman filter.  
We start by deriving the dual result (in the sense of \cite{Kalman_Bucy:1961,Kalman:1960}) of the work \cite{Bamieh:2002}. By utilizing the spatial Fourier transform,  we show that the infinite-dimensional algebraic Riccati equation (ARE) for the filter \textit{decouples} into an infinite family of \textit{finite-dimensional} AREs parametrized by the spatial frequency.  Such a decoupling eases solving the infinite-dimensional ARE and allows us to obtain its  solution explicitly. It follows that the Kalman gain operator is a \textit{spatial convolution}. In addition,  studying the region of analyticity of the extension of the Fourier symbol of the gain operator shows that the convolution kernel is \textit{spatially localized}.  Thus,  the filter is spatially invariant itself and exhibits an inherent degree of spatial localization.

\paragraph*{Main Contributions }
\indent \textit{(1)} We characterize the 
spatial decay rate of the Kalman gain and establish how the interplay between the Fourier symbol of the operator in the plant dynamics and the power spectral densities of process and measurement noises determines the information structures of the filter. We find that noise variances and their spatial autocorrelations 
 affect the degree of spatial localization of the Kalman gain.
 We provide a \textit{matching condition} under which the Kalman filter becomes completely decentralized.  
\textit{(2)} In analogy with the root locus we introduce a new graphical technique, that we name the \textit{Branch Point Locus}, to determine the spatial localization of the Kalman gain by inspection of the trajectories of its branch points in the complex spatial frequency plane. 
 \textit{(3)} We illustrate our analysis with two case studies: Kalman filters for \textit{i)} diffusion (extending the results presented by the authors in \cite{Arbelaiz:2020}),  and \textit{ii)} the linearized Swift-Hohenberg equation, both over the real line.  We use these examples to highlight the usefulness of \textit{dimensional analysis}  to reduce the dimensionality of the parameter space and provide physical interpretations of our results.


\paragraph*{Paper Structure} Mathematical background and notation are in Section \ref{sec:Mathematical_pre}. SIS and the assumptions we follow throughout this work are introduced in Section \ref{subsec:the_plant}. Section \ref{subsec:the_filtering_problem}  formulates the spatiotemporal filtering problem for SIS and Section \ref{subsec:main_results} summarizes our main findings. To streamline the presentation, technical results and background regarding structural properties of the Kalman filter for SIS are relegated to Appendix \hyperref[sec:Appendix_II]{I}.
Section \ref{sec:information_structures_general} analyzes the spatial locality of the Kalman gain  and provides a condition --- the ``matching" condition --- under which the filter is completely decentralized.  Section \ref{sec:spatial_llocalization_and_performance} considers plants with dynamics governed by differential operators of even order.  For these,  we explicitly characterize the asymptotic spatial decay rate of the Kalman gain when the plant is subject to spatiotemporal white noises.  
Case studies illustrating our theoretical results are presented in Section \ref{sec:case_studies}.  We draw conclusions in Section \ref{sec:conclusion}. 




\section{Mathematical Preliminaries}
\label{sec:Mathematical_pre}

\paragraph{Notation} 
$\mathbb{N}$ denotes the set of positive integers. 
 Lower case letters $\psi(x,t)$ denote the value of a scalar \textit{spatiotemporal field}\footnote{We use the term \textit{scalar spatiotemporal field} to denote a function associating a single number to each space-time coordinate in its domain.}  at location $x \in \mathbb{R}$ and time $t\in \mathbb{R}_{\geq 0}$.  Bold font denotes $\boldsymbol{\psi}(t) :=  \psi(\cdot, t) $,  a spatially distributed signal at time $t$.  Sometimes  we define dimensionless variables \cite{IsaakovichBarenblatt2006}.  If the dimensional variable is denoted by Greek (roman) font,  we use capital Greek (san-serif) font to denote the dimensionless counterpart; e.g.,  if $\psi$  is a dimensional field,  $\Psi$ denotes its dimensionless analogue.  We are mostly interested in signals $\boldsymbol{\psi}(t)\in L^2(\mathbb{R})$ -- the Hilbert space of equivalence classes of square-integrable functions -- equipped with the inner product 
\begin{equation}
\langle f, g \rangle_{L^2(\mathbb{R})} \coloneqq \int_{\mathbb{R}} f(x)g(x)^* \mathrm{d}x,
\label{eq:L2_inner_product}
\end{equation}
where $(^*)$ denotes the complex conjugate of a complex number.   Sometimes,  we write $f \in L^2_{\mathbb{C}}(\mathbb{R})$ to emphasize that $f$ is complex-valued and denote its real and imaginary parts by $\Re(f)$ and $\Im(f)$, respectively. The imaginary unit is denoted by $\mathbf{i}:= \sqrt{-1}$.  We use $\langle f,  g \rangle$ to denote the inner product \eqref{eq:L2_inner_product} and $\|\boldsymbol{\cdot} \|$ to denote the induced norm in $L^2(\mathbb{R})$,  that is,  $\|f\| = \|f\|_{L^2(\mathbb{R})} := \langle  f,f \rangle^{\frac{1}{2}}$. We use subindexes to distinguish inner products and norms in spaces different from $L^2$ when needed.  The absolute value of a complex number is denoted by $|\boldsymbol{\cdot}|$. 
We denote by $H_k(\mathbb{R})$ the set $H_k(\mathbb{R}) := \big\{ f \in L^2(\mathbb{R}): \partial^{\alpha} f \in L^2(\mathbb{R}) \text{ for all }\alpha \leq k\big\},$ with $\alpha$ and $k$ non-negative integers. 
This is the Sobolev space of order $k$.  
Its image under the Fourier transform --- see Def. \ref{def:fourier_transform}  --- is $\hat{H}_k(\mathbb{R}) = \{\hat{f} \in L^2(\mathbb{R}): |\lambda|^\alpha \hat{f}(\lambda) \in L^2(\mathbb{R}) \}.$
We use caligraphic font to denote a linear operator  $\mathcal{A}: \mathcal{D}(\mathcal{A}) \subset L^2(\mathbb{R}) \to L^2(\mathbb{R})$, where $\mathcal{D}(\mathcal{A})$ denotes the domain of $\mathcal{A}$. 
We overload notation and use $\mathcal{A}^*$ to denote the adjoint of a linear operator $\mathcal{A}$; if a kernel representation of $\mathcal{A}$ exists \cite[Ch. 15]{Ascoli1977,Duistermaat2010}, we use  
$A: \mathbb{R} \times \mathbb{R} \to \mathbb{R}$ to denote the kernel 
\begin{subequations}
\begin{equation}
[\mathcal{A}f](x) \coloneqq \int_{\mathbb{R}} A(x,\xi) f(\xi)  \mathrm{d}\xi.
\label{eq:kernel_representation_operator} 
\end{equation}
If the kernel is a function of a single variable,  we refer to it as a \textit{Toeplitz kernel} and it defines a \textit{convolution operator}
\begin{equation}
[\mathcal{A}f](x) \coloneqq \int_{\mathbb{R}} A(x-\xi) f(\xi) \mathrm{d}\xi.
\label{eq:convolution_operator_def}
\end{equation}
\label{eq:integral_representations}
\end{subequations}

\paragraph{Spatial Invariance and Fourier Analysis}
\begin{defn}[Translation operator,  $\mathcal{T}_z$] \label{defn:traslation_operator} The translation (or shift) operator denoted by $\mathcal{T}_z:L^2(\mathbb{R}) \rightarrow L^2(\mathbb{R})$ is defined as $[T_z f](x)\coloneqq f(x-z)$ for any $z\in \mathbb{R}$ and $f \in L^2(\mathbb{R})$.  
\end{defn}

\begin{defn}[Translation invariant operator, from \cite{Bamieh:2002}] \label{defn:traslation_invariant_operator}
An operator $\mathcal{A}$ is translation invariant if $\mathcal{T}_z: \mathcal{D}(\mathcal{A}) \rightarrow \mathcal{D}(\mathcal{A})$ and $\mathcal{A} \mathcal{T}_z = \mathcal{T}_z \mathcal{A}$,  for every translation $\mathcal{T}_z$.  Since this work focuses on translation invariance in the spatial coordinate, we use the terms \textit{translation invariance} and \textit{spatial invariance} interchangeably.   Spatial convolutions \eqref{eq:convolution_operator_def} are spatially invariant operators.  
\end{defn}

\begin{defn}[Multiplication operator] 
\label{defn:multiplication_operator}
A multiplication operator $\mathcal{M}_A: \mathcal{D}(\mathcal{M}_A) \subseteq L^2(\mathbb{R}) \to L^2(\mathbb{R})$ is defined by $[\mathcal{M}_A f](x) := A(x)f(x)$. 
We refer to $A$ as the \textit{symbol} of the multiplication operator $\mathcal{M}_A$.
\end{defn}

\begin{defn}[spatial Fourier transform]
\label{def:fourier_transform}
Let $f$ denote a spatiotemporal field with spatial coordinate $x \in \mathbb{R}$ and $t \in \mathbb{R}_{\geq 0}$.
The \textit{spatial} Fourier transform 
of $f$ is
\begin{align}
    \hat{f}(\lambda,t) & \coloneqq \frac{1}{\sqrt{2\pi}} \int_{\mathbb{R}} f(x,t) e^{-\mathbf{i} \lambda x}\mathrm{d}x, \label{eq:spatial_FT}
\end{align}
where $\lambda \in \mathbb{R}$ denotes the spatial frequency and $\mathbf{i}$ the imaginary unit.  Sometimes we denote the spatial Fourier transform by $\mathcal{F}(\boldsymbol{\cdot})$.  We indicate parametrization by $\lambda$ using $\hat{f}(\lambda,t)$ or $\hat{f}_{\lambda}(t)$  interchangeably. 
\end{defn} 
Let $f(\cdot,t) \in L^2(\mathbb{R})$.  The \textit{Parseval-Plancherel identity} with the Fourier transform normalization defined in \eqref{eq:spatial_FT} is:
\begin{equation}
\int_{\mathbb{R}} |f(x,t)|^2 \mathrm{d}x = \int_{\mathbb{R}} |\hat{f}_{\lambda}(t)|^2 \mathrm{d}\lambda.
    \label{eq:plancherel}
\end{equation}
More generally, $\langle f, g \rangle = \langle \hat{f}, \hat{g} \rangle$ for $f,g \in L^2(\mathbb{R})$. 

The spatial Fourier transform \eqref{eq:spatial_FT} \textit{diagonalizes} spatially invariant operators, in the sense that they are transformed into multiplication operators in the spatial frequency domain.
For a spatially invariant operator $\mathcal{A}$, $\mathcal{F}[\mathcal{A}h](\lambda) = \hat{\mathcal{A}}(\lambda) \hat{h}(\lambda)$. Following Definition \ref{defn:multiplication_operator}, we refer to 
$\hat{\mathcal{A}}_{\lambda}: \mathbb{R} \to \mathbb{C}$ as the \textit{Fourier symbol} \cite{Bamieh:2002} of the spatially invariant operator $\mathcal{A}$.  

For a densely defined linear spatially invariant operator $\mathcal{A}: \mathcal{D}(\mathcal{A}) \subset L^2(\mathbb{R}) \to  L^2(\mathbb{R})$ with Fourier symbol $\hat{\mathcal{A}}_{\lambda}$ and kernel representation \eqref{eq:convolution_operator_def}, $\hat{\mathcal{A}}_{\lambda}$ and $\hat{A}_{\lambda}$ are related by $\hat{\mathcal{A}}_{\lambda}  = \sqrt{2 \pi} \, \hat{A}_{\lambda}$ and the Fourier symbols of the adjoint and inverse operators are
$\widehat{\mathcal{A}_{\lambda}^*}  = \hat{\mathcal{A}}^*_{\lambda}$
and 
$\widehat{\mathcal{A}^{-1}_{\lambda}}  = \hat{\mathcal{A}}^{-1}_{\lambda}$, respectively.

\begin{defn}[Extension to the complex plane] \label{defn:extension}
Given a function $\hat{\mathcal{L}}_{\lambda}: \mathbb{R} \to \mathbb{C}$, 
we construct its extension to the complex plane,  $\hat{\mathcal{L}}_{z}: \mathbb{C} \to \mathbb{C}$,  by replacing $\lambda$ in $\hat{\mathcal{L}}_{\lambda}$ by $(-\mathbf{i}z)$  with $z\in \mathbb{C}$.
\end{defn}

In this work, we use Definition \ref{defn:extension} to define the extension of Fourier symbols of  multiplication operators to the complex plane.

\begin{thm}[Paley-Wiener Theorem,  from \cite{Bamieh:2002} originally adapted from \cite{Hormander:1990}] \label{thm:Paley-Wiener}
Let $\Gamma$ be an open interval in $\mathbb{R}$. Let the extension $\hat{\mathcal{L}}_z$ be 
analytic 
on the strip $\Gamma+ \mathbf{i} \mathbb{R}$ and such that for every compact set $\Gamma_0 \subset \Gamma$, there exists $C, N>0$ for which
\begin{equation}
|\hat{\mathcal{L}}_z| \leq C(1+|z|)^N
\label{eq:polynomial_bound}
\end{equation}
holds $\forall \, \Re(z) \in \Gamma_0$. Then,  there exists a 
distribution $\mathcal{L}$ such that $e^{-\eta x} \mathcal{L}$ is a tempered distribution for every $\eta \in \Gamma$. 
\end{thm}

\section{Problem formulation \& Main results}
\label{sec:Problem_Formulation}
 In this section,  we introduce the spatially invariant system of interest and formulate the optimal spatiotemporal filtering problem. We also summarize our main findings.
To streamline the exposition,  technical results regarding structural properties of the optimal filter are relegated to Appendix \hyperref[sec:Appendix_II]{I}.

\subsection{The spatially invariant system}
\label{subsec:the_plant}
We study spatially distributed systems with spatiotemporal linear dynamics in continuous time.  The system is described in 
state-space representation \cite{Bala:1981}\footnote{In the plant \eqref{eq:plant_abstract} we consider a more general $\mathcal{G}$ than the one provided in the filtering problem formulation of \cite{Bala:1981}.} as:
\begin{subequations}
\begin{align}
\frac{\mathrm{d} \boldsymbol{\psi}}{\mathrm{d} t} (t)& = \mathcal{A} \boldsymbol{\psi}(t) + \mathcal{B} \boldsymbol{w}(t),  \label{eq:dynamics_abstract}\\
\boldsymbol{y}(t) & = \mathcal{C} \boldsymbol{\psi}(t) + \mathcal{G} \boldsymbol{v}(t),  \label{eq:measurement_abstract}
\end{align}
\label{eq:plant_abstract}
\end{subequations}
with $t \in \mathbb{R}_{\geq 0}$ the instant of time.   $\boldsymbol{\psi}(t), \boldsymbol{w}(t)$, and $\boldsymbol{v}(t)$ 
are Hilbert space-valued random variables.
We refer to $\boldsymbol{\psi}(t)$ as the \textit{state}. The measurement $\boldsymbol{y}$ is distributed in space and corrupted by noise $\boldsymbol{v}$.  Dynamics are subject to the noise process $\boldsymbol{w}$.  
The random initial condition $\boldsymbol{\psi}(0)$ 
follows a standard Gaussian, independent from process and measurement noises. Throughout this work,  the following assumptions hold: 


\begin{assumption}[Scalar and real spatiotemporal fields]
\label{ass:A1}
    The spatiotemporal signal $\psi$,  noises $w$ and $v$,  and measurement $y$ are real scalar spatiotemporal random fields \cite{Christakos2017}. 
\end{assumption}

\begin{assumption}[White noise processes]
\label{ass:A2}
    $\boldsymbol{w}$ and $\boldsymbol{v}$ are independent white noises as defined in \cite[Section 6.6]{Bala:1981}: zero-mean Gaussian with identity covariance operators.
\end{assumption}

\begin{assumption}[The operator $\mathcal{A}$] 
\label{ass:A3}
    The operator $\mathcal{A}: \mathcal{D}(\mathcal{A}) \subset L^2(\mathbb{R}) \rightarrow L^2(\mathbb{R})$ is a densely defined differential operator  
and the infinitesimal generator of a strongly continuous ($C_0$) 
 semigroup in $L^2(\mathbb{R})$, with continuous Fourier symbol $\hat{\mathcal{A}}_{\lambda}$.
\end{assumption}

\begin{assumption}[The operators $\mathcal{B}, \mathcal{C}, \& \; \mathcal{G}$] \hfill
\label{ass:operators_BCG}
\begin{enumerate}
    \item $\mathcal{B}: L^2(\mathbb{R}) \to L^2(\mathbb{R})$,    $\mathcal{G}: L^2(\mathbb{R}) \to L^2(\mathbb{R})$ and $\mathcal{C}: \mathcal{D}(\mathcal{C}) \subseteq L^2(\mathbb{R}) \to L^2(\mathbb{R})$ are linear  
    time independent  spatially invariant operators.
    \item  Their respective Fourier symbols $\hat{\mathcal{B}}_{\lambda}: \mathbb{R} \to \mathbb{C}$,   $\hat{\mathcal{G}}_{\lambda}: \mathbb{R} \to \mathbb{C}$ and $\hat{\mathcal{C}}_{\lambda}: \mathbb{R} \to \mathbb{C}$ 
are
proper rational functions of $\lambda$, without zeros nor real poles. 

\item $(\mathcal{G} \mathcal{G}^*)$
is a positive definite invertible operator. $(\mathcal{B} \mathcal{B}^*)$
is a positive definite operator.
\end{enumerate}
\end{assumption}
The spatial operators $\mathcal{B}$ and $\mathcal{G}$ are interpreted as \textit{spatial shaping filters}.  
We refer to $\widehat{\mathcal{B}_{\lambda} \mathcal{B}_{\lambda}^*}$ and $\widehat{\mathcal{G}_{\lambda} \mathcal{G}_{\lambda}^*}$ as the \textit{power spectral densities} of process and measurement noises,  respectively.  

\begin{assumption}
\label{ass:5}
    The pairs $(\mathcal{A},\mathcal{B})$ and $(\mathcal{A}^*, \mathcal{C}^*)$ are exponentially stabilizable. 
\end{assumption}



Since by Assumption \ref{ass:A2}  the noise fields are spatiotemporally white,  they are stationary.
By Assumptions \ref{ass:A3} and \ref{ass:operators_BCG}.1, the operators in the state-evolution equation and measurement are spatially invariant.  
Hence,  all the operators in the system are spatially invariant and we refer to \eqref{eq:plant_abstract} as a \textit{spatially invariant system}. In this work, we leverage such spatially invariant structures to gain analytical tractability in the design and characterization of the optimal filter. Assumptions \ref{ass:operators_BCG} and \ref{ass:5} ensure the well-posedness of the filtering problem \cite{Zabczyk1975} -- i.e., the existence and uniqueness of the stabilizing solution of
the Riccati equation corresponding to \eqref{eq:plant_abstract}.


\subsection{The Filtering Problem}
\label{subsec:the_filtering_problem}

We aim to design a temporally causal 
estimator $\boldsymbol{\tilde{\psi}}(t)$
of the state $\boldsymbol{\psi}(t)$ minimizer of the steady-state variance of the estimation error $\mathbf{e}$,  defined as 
$
\mathbf{e}(t) \coloneqq \boldsymbol{\psi}(t) - \boldsymbol{\tilde{\psi}}(t).
\label{eq:estimation_error_def}
$
The optimal state estimate is the conditional expectation $\boldsymbol{\tilde{\psi}}(t) = \mathbb{E}\big[\boldsymbol{\psi}(t)| \{\boldsymbol{y}(\tau); \tau \leq t\}\big]$.  Its dynamics are described by the \textit{distributed-parameter Kalman filter} \cite{Bala:1981} 
\begin{equation}
\frac{\mathrm{d} \boldsymbol{\tilde{\psi}}}{\mathrm{d} t}(t) = (\mathcal{A} - \mathcal{L} \mathcal{C})\boldsymbol{\tilde{\psi}}(t) + \mathcal{L} \, \boldsymbol{y}(t),  \label{eq:KBF_abstract_form}
\end{equation}
where 
$\boldsymbol{\tilde{\psi}}(0) = \boldsymbol{0}$ and the feedback operator  is
$
\mathcal{L} = \mathcal{P} \mathcal{C}^* (\mathcal{G} \mathcal{G}^*)^{-1}.
\label{eq:feedback_operator}
$\
In analogy with the finite dimensional setting,  we refer to  $\mathcal{L}$ as the \textit{Kalman gain}. $\mathcal{P}$ denotes the steady-state covariance operator of the optimal estimation error and satisfies the algebraic Riccati equation \eqref{eq:estimation_OARE}.

 In our spatially invariant setting, the operators $\mathcal{P}$ and $\mathcal{L}$ are spatially invariant themselves.  Their respective Fourier symbols satisfy \eqref{eq:ARE_frequency_almost_everywhere_theorem} and are\footnote{
Technical details on the spatial invariance of $\mathcal{P}$ and $\mathcal{L}$, together with the computation of their respective Fourier symbols 
are provided in Appendix \hyperref[subsec:structural_properties_KF]{I-B}. 
}
\begin{subequations}
\begin{align}
 \hat{\mathcal{P}}_{\lambda} & = \frac{\Re(\hat{\mathcal{A}}_{\lambda}) |\hat{\mathcal{G}}_{\lambda}|^2}{|\hat{\mathcal{C}}_{\lambda}|^2} + \sqrt{\frac{\Re(\hat{\mathcal{A}}_{\lambda})^2 |\hat{\mathcal{G}}_{\lambda}|^4}{|\hat{\mathcal{C}}_{\lambda}|^4} + \frac{|\hat{\mathcal{B}}_{\lambda}|^2 |\hat{\mathcal{G}}_{\lambda}|^2}{|\hat{\mathcal{C}}_{\lambda}|^2}},
 \label{eq:Fourier_symbol_covariance}\\
 \hat{\mathcal{L}}_{\lambda} & = \frac{\Re(\hat{\mathcal{A}}_{\lambda})}{\hat{\mathcal{C}}_{\lambda}} + \sqrt{\frac{\Re(\hat{\mathcal{A}}_{\lambda})^2}{\hat{\mathcal{C}}_{\lambda}^2} + \frac{|\hat{\mathcal{B}}_{\lambda}|^2\hat{\mathcal{C}}_{\lambda}^*}{|\hat{\mathcal{G}}_{\lambda}|^2 \hat{\mathcal{C}}_{\lambda}}}.  
\label{eq:Fourier_symbol_KalmanGain_bis}
\end{align}
\label{eq:P_L_symbols}
\end{subequations}
In the physical domain, these correspond to \textit{spatial convolutions}.

\begin{assumption}[Full state noisy measurements]
\label{ass:A4}
     We are interested in the setting in which measurements are homogeneously distributed in the spatial domain and provide noisy observations of the full state, that is, $\boldsymbol{y}(t) = \boldsymbol{\psi}(t) + \mathcal{G}\boldsymbol{v}(t)$. \end{assumption}

\subsection{Main Results}
\label{subsec:main_results}
In this work, we characterize how valuable measurements from far away are for filtering (i.e., the degree of spatial locality of the filter) as a function of the variances and spatial autocorrelations of the noises perturbing the plant.

We connect how the spatial decay rate of the optimal filter relates to the branch points of the analytic continuation of the Fourier symbol $\hat{\mathcal{L}}_{\lambda}$. Using this relationship we define a new graphical technique, which we name the \textit{Branch Point Locus} (BPL), to analyze the spatial locality of the optimal filter by inspecting analyticity regions in the complex \textit{spatial} frequency plane. We illustrate how dimensional analysis is useful in this task, as it allows us to reduce the dimensionality of the parameter space of the plant.

We start by deriving a \textit{matching condition} under which the optimal filter is completely decentralized: 

\begin{prop*}
Let $\mathcal{A}$, $\mathcal{B}$, $\mathcal{G}$ be given in \eqref{eq:plant_abstract} and let Assumptions \ref{ass:A1} to \ref{ass:A4} hold.  If there exists a constant $\ell \in \mathbb{R}_+$ such that the relationship
$
|\hat{\mathcal{B}}_{\lambda}|^2
= \ell\big(\ell - 2 \Re(\hat{\mathcal{A}}_{\lambda})\big) |\hat{\mathcal{G}}_{\lambda}|^2
\label{eq:decentralization_condition}
$
is satisfied
for all $\lambda \in \mathbb{R}$, then the Kalman filter \eqref{eq:KBF_abstract_form} is completely decentralized.  
\end{prop*}

When such a \textit{matching condition} is met, collecting neighboring measurements does not improve the filtering performance. This is a setting of extreme spatial localization: at each spatial site $x$ only the measurement at $x$ is needed for filtering. In our problem set-up, the matching condition is typically met when measurement noise is spatially autocorrelated and the autocorrelation lengthscale matches the characteristic lengthscale of the dynamics. This result highlights the importance of accounting for the spatial autocorrelations of the noise  in the filter synthesis. 

We also analyze the common setting in which $\mathcal{A}$ is a differential operator with Fourier symbol $\hat{\mathcal{A}}_{\lambda} = -|a|\lambda^{2n}, n\in \mathbb{N}$. We explicitly characterize the degree of spatial locality of the corresponding optimal filter:

\begin{thm*}[informal]
Let Assumptions \ref{ass:A1} and \ref{ass:A2} hold and let the Fourier symbols in the spatially invariant system \eqref{eq:plant_abstract} be $\hat{\mathcal{A}}_{\lambda} = -|a|\lambda^{2n}$, $\hat{\mathcal{B}}_{\lambda}=\sigma_w$, $\hat{\mathcal{G}}_{\lambda} = \sigma_v$ and $\hat{\mathcal{C}}_{\lambda} = 1$  with $\sigma_v, \sigma_w >0$. Then, the Kalman filter exhibits asymptotic exponential spatial decay, with decay rate\footnote{We formalize what is meant by spatial decay of the filter in Section \ref{sec:information_structures_general} and in Appendix \hyperref[subsec:spatial_localization_KF_appendix]{I-C}. } proportional to
$ 
\big( \frac{\sigma_w}{|a| \, \sigma_v}\big)^{\frac{1}{2n}}.
\label{eq:monomials_ratio}
$
\end{thm*}

The implication of such a spatial decay is that the value of measurements for filtering 
at a fixed spatial site $x$ decays fast with the distance from $x$ to the sensor, that is, the further away the sensor, the less valuable the measurement for filtering at $x$. Furthermore, the theorem establishes that if measurements are of good quality (i.e., small sensor noise variance, low $\sigma_v^2$) or process noise is large (i.e., large uncertainty in the model of the dynamics, high $\sigma_w^2$), then the spatial localization of the filter increases, making it amenable to decentralized implementations.




\section{Spatial locality of the Kalman filter for SIS}
\label{sec:information_structures_general}

\begin{figure}[H]
\centering
\includegraphics[width=0.895\linewidth]{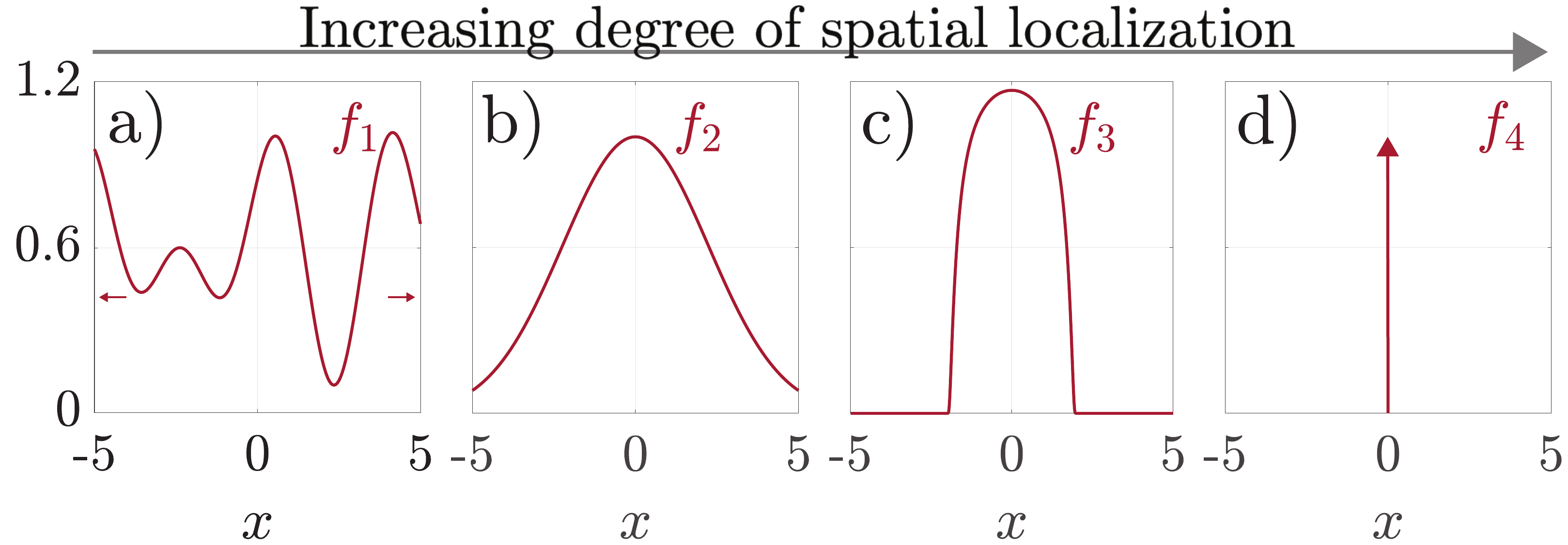}
\caption{ \small{Notions of spatial locality.  a) $f_1$ is a \textit{spatially distributed} signal.  b) $f_2$ is a Gaussian centered at the origin: spatially distributed,  but \textit{localized} (rapidly decaying).  c) $f_3$ is a bump function,  
with compact support
 --- an example of \textit{decentralized} or \textit{space-limited} kernel.  d) $f_4$ is a Dirac delta distribution, point-supported and thus,  \textit{completely decentralized}.  Vertical axis is the same in all panels.}}
\label{fig:localization_kernels}
\end{figure}

\paragraph{Defining locality in space} We say that a spatial signal is \textit{distributed} when it covers the entire spatial domain (Fig.  \ref{fig:localization_kernels}a); e.g.,  in our setting,  ``distributed measurements" refers to measurements being collected from the whole real line.  
We are interested in characterizing the information structures (that is,  the spatial locality) of a spatial feedback operator.  We characterize locality by analyzing the spatial decay properties of the corresponding kernel in the integral representation \eqref{eq:integral_representations}. 
We say that the feedback operator is \textit{spatially localized} if its kernel is spatially distributed,  but concentrated \cite{Slepian:1983} around its center (Fig.  \ref{fig:localization_kernels}b);  when its kernel is compactly supported,  we say the feedback is \textit{decentralized} or \textit{space-limited} (Fig.  \ref{fig:localization_kernels}c); and if the kernel is point-supported,  we say the feedback is \textit{completely decentralized} (Fig.  \ref{fig:localization_kernels}d).  
\paragraph{Spatial locality in the Kalman filter} Measurements $\boldsymbol{y}(t)$ only enter the Kalman filter   \eqref{eq:KBF_abstract_form}  through the Kalman gain $\mathcal{L}$, which is a spatial convolution in our problem set-up (see Theorem \ref{thm:KB_freq}).  State estimates $\tilde{\boldsymbol{\psi}}(t)$ also need to be exchanged within different spatial sites,  but since by Assumption \ref{ass:A3} $\mathcal{A}$ is a differential operator 
and by Assumption \ref{ass:A4} the filter has access to full state measurements,  $\mathcal{L}$ sets the degree of spatial localization of the closed-loop.   Thus,  characterizing
the information exchange required
among different spatial locations in the system for filtering --- the \textit{information structures} of the filter --- amounts to characterizing the spatial behavior of the convolution operator $\mathcal{L}$. 
If the kernel $L$ of the operator $\mathcal{L}$ is \textit{compactly supported},  the filter is \textit{decentralized} and only requires local measurements (e.g.,  Fig. \ref{fig:localization_kernels}c). The  degree of decentralization is defined by $\mathrm{supp}(L)$.  Point-supported kernels yield complete decentralization
(e.g.,  Fig. \ref{fig:localization_kernels}d).
Alternatively,  if $L$ decays in space,  its \textit{spatial decay rate} dictates the relevance of measurements for the filter as a function of distance to the sensor where the measurement is collected; we refer to this setting as \textit{centralized but localized} (e.g.,  Fig. \ref{fig:localization_kernels}b) as measurements from the whole spatial domain need to be communicated for the filtering procedure at each spatial site. The degree of spatial localization is dictated by the spatial decay rate of $L$.
From \eqref{eq:Fourier_symbol_KalmanGain_bis}, 
we deduce that $L$ will not be compactly supported (i.e.,  decentralized) in general, as due to the square root in $\hat{\mathcal{L}}_{\lambda}$  its extension $\hat{\mathcal{L}}_z$ to the complex plane (see Def. \ref{defn:extension}) is not an entire analytic function.
This implies that in general, for a fixed spatial location $x$,  measurements from the whole spatial domain are needed to compute the optimal state estimate at $x$.

 The LQR control gain for SIS is spatially localized \cite{Bamieh:2002}. Similarly, in our filtering problem, so is the Kalman gain $\mathcal{L}$ (see Appendix \hyperref[subsec:spatial_localization_KF_appendix]{II-C} for technical details).
Particularly for our problem set-up, by Assumptions \ref{ass:A3} and \ref{ass:operators_BCG}$,  \hat{\mathcal{A}}_{\lambda}$ is a polynomial  and $\hat{\mathcal{G}}_{\lambda}$ and $\hat{\mathcal{B}}_{\lambda}$ are proper rational functions, respectively.  Hence,  
the Kalman gain $\mathcal{L}$ might contain point-supported distributions at the origin 
 and a regular component 
that asymptotically ($|x| \to \infty$) decays to zero faster than any exponential $e^{-|\eta||x|}$ with $|\eta|<\theta$ and with $\theta$ as defined in \eqref{eq:theta_def}. 
For simplicity,  from this point onwards we refer to $\theta$ as the \textit{asymptotic (exponential) spatial decay rate of $\mathcal{L}$}.
The rapid spatial decay of the operator $\mathcal{L}$ implies that even in a \textit{centralized} setting,  the Kalman filter exhibits an \textit{inherent degree of spatial localization}.  Informally,  this means that given a spatial location $x$,  measurements from its neighborhood are more heavily weighted by the filter than measurements from far away.  
Such spatial localization motivates the idea of designing a decentralized filter architecture 
by \textit{spatial truncation} of the tails of the kernel $L$ beyond a desired distance $T$  \cite{Bamieh:2002}:
$
L_T(x) \coloneqq
\begin{cases}
L(x), \;\; \text{if} \;\; |x|\leq T,\\
0, \;\;\;\;\;\;\;\, \text{if} \;\; |x|>T.
\end{cases}
\label{eq:LT_spatial_truncation}
$
However,  such an $L_T$ will not be optimal in general and might even lead to instability of the filter dynamics \cite{Motee:2008}. 
Based on the results introduced in \cite{Arbelaiz2021},  in a follow-up paper \cite{Arbelaiz2022_partII} we propose a design procedure to synthesize the optimal stabilizing compactly supported $L_T$ through a convex functional optimization. However, prior to solving this problem, 
it is useful to determine the spatial decay rate of the centralized filter as a function of system parameters in order to select a reasonable value for $T$. Such characterization is the aim of this work.
We show that the spatial decay depends on how the power of the noise processes perturbing the plant is allocated in the different spatial frequencies $\lambda$ after the  spatial shaping performed by $\mathcal{B}$ and $\mathcal{G}$ (i.e., on the variances and spatial autocorrelations introduced by $\mathcal{B}$ and $\mathcal{G}$).
We start by deriving a condition under which the Kalman filter is completely decentralized.

  

\subsection{A sufficient condition for complete decentralization}
\label{subsec:complete_decentralization}

\begin{defn}[Completely decentralized Kalman filter]
\label{defn:completely_decentralized}
Under Assumptions \ref{ass:A1} to \ref{ass:A4}, we say that the Kalman filter \eqref{eq:KBF_abstract_form} for the spatially invariant plant \eqref{eq:plant_abstract}  is \textit{completely decentralized} if it is of the form 
\begin{equation}
\frac{\mathrm{d} \boldsymbol{\tilde{\psi}}}{\mathrm{d} t}(t) = (\mathcal{A}-\ell \, \mathcal{I}) \, \boldsymbol{\tilde{\psi}}(t) \; + \ell \, \boldsymbol{y}(t),
\label{eq:KBF_totally_decentralized_PDE_def}
\end{equation}
where $\ell$ is a positive real constant.   
A similar definition of decentralized operator was introduced in \cite{DeCastro2002}.
\end{defn}

The filter \eqref{eq:KBF_totally_decentralized_PDE_def} is completely decentralized in the sense that for a fixed spatial location $x$,  it only requires the measurement from $x$ to be implemented.
When such a filter is optimal, accessing measurements from neighboring spatial locations does not improve performance.  We show that if a matching condition is met among the Fourier symbols of the operators in the plant, then the Kalman filter takes the form \eqref{eq:KBF_totally_decentralized_PDE_def}.  

\begin{prop}[Condition for complete decentralization] 
\label{prop:complete_decentralization}
For given $\mathcal{A}$, $\mathcal{B}$ and $\mathcal{G}$ in \eqref{eq:plant_abstract} and under Assumptions \ref{ass:A1} to \ref{ass:A4},  if there exists a constant $\ell \in \mathbb{R}_+$ such that the relationship 
\begin{equation}
|\hat{\mathcal{B}}_{\lambda}|^2
= \ell\big(\ell - 2 \Re(\hat{\mathcal{A}}_{\lambda})\big) |\hat{\mathcal{G}}_{\lambda}|^2
\label{eq:decentralization_condition}
\end{equation}
is satisfied for all $\lambda \in \mathbb{R}$, then the Kalman filter \eqref{eq:KBF_abstract_form} is completely decentralized.  
We refer to \eqref{eq:decentralization_condition} as the matching condition.  
\end{prop}
\begin{proof}
Since Assumptions \ref{ass:A1} to \ref{ass:5} hold, the system  \eqref{eq:plant_abstract} is spatially invariant, the Kalman filter \eqref{eq:KBF_abstract_form} is well-posed, and the Fourier symbol of its Kalman gain is given by \eqref{eq:Fourier_symbol_KalmanGain_bis}.  Substitute Assumption \ref{ass:A4} and the matching condition \eqref{eq:decentralization_condition} in \eqref{eq:Fourier_symbol_KalmanGain_bis} to obtain $\hat{\mathcal{L}}_{\lambda} = \Re(\hat{\mathcal{A}}_{\lambda}) + \sqrt{\big(\ell - \Re(\hat{\mathcal{A}}_{\lambda})\big)^2} = \ell$.  Taking the inverse spatial Fourier transform yields a filter of the form \eqref{eq:KBF_totally_decentralized_PDE_def}. 
\end{proof}

Hence,  
the matching condition will typically hold when $\mathcal{G}$ introduces short-range (exponential) spatial autocorrelations in the measurement noise and parameters of the plant ``match" (we illustrate the matching through a case study in Section \ref{sec:case_studies}).
This observation highlights the important role that spatial autocorrelations of the noise processes play in defining the information structures of the Kalman filter (i.e., the usefulness of far-away measurements for estimation).  
However,  the effect of spatial autocorrelations on optimal estimators is usually disregarded in the literature \cite{Liu2016},  as noises are often assumed spatially uncorrelated.   Autocorrelated noise, nevertheless, is of practical importance 
\cite{Ucinski:2020}.  
We characterize the effect that noise statistics (variances and spatial autocorrelations) have on the information structures of Kalman filters for SIS and on their performance.
We start by doing so in the setting in which the operator $\mathcal{A}$ is a differential operator of even order $2n$ ($n \in \mathbb{N}$).  To the best of our knowledge, this is the first time that the 
spatial locality of the Kalman gain is explicitly characterized as a function of the noise processes in the system.

\section{Differential Operators of Even Order: \\ information structures \& performance}
\label{sec:spatial_llocalization_and_performance}
We analyze the Kalman filter for a plant of the form 
\begin{subequations}
\begin{align}
\frac{d \boldsymbol{\psi}}{d t} (t) & = \mathcal{A} \boldsymbol{\psi}(t) + \sigma_w \boldsymbol{w}(t),  \label{eq:dynamics_monomial}\\
\boldsymbol{y}(t) & =  \boldsymbol{\psi} (t) + \sigma_v \boldsymbol{v}(t),
\label{eq:measurement_monomial}
\end{align}
\label{eq:plant_monomial}
\end{subequations}
where the 
operator 
$
\mathcal{A}\coloneqq a \, \partial_x^{(2n)}
$
with $\mathcal{D}(\mathcal{A}) = H_{(2n)}(\mathbb{R})$, $n \in \mathbb{N}$,
and $a>0 \; (a<0)$ when $n$ is odd (even); $ t \in \mathbb{R}_{\geq 0}$, and $\sigma_w, \sigma_v >0$;   
$\boldsymbol{w}$ and $\boldsymbol{v}$ satisfy Assumption \ref{ass:A2}.
The operator $\mathcal{A}$ 
is spatially invariant with Fourier symbol $\hat{\mathcal{A}}_{\lambda} = -|a| \,  \lambda^{2n}$, which satisfies the half-plane condition \eqref{eq:half_plane_condition}.  Then,  by Proposition \ref{prop:C0_generation} $\mathcal{A}$ is the infinitesimal generator of a $C_0$-semigroup in $L^2(\mathbb{R})$.
Consequently,  we use Theorem \ref{thm:KB_freq}  to study the distributed-parameter Kalman filter for the plant \eqref{eq:plant_monomial} in the spatial frequency domain.  We explicitly characterize the scalings of the asymptotic spatial decay rate of the Kalman gain $\mathcal{L}$ and the performance of the filter on plant's parameters.




\subsection{Information structures}
\label{subsec:Information Structures}

\begin{thm}[Asymptotic spatial decay rate of $\mathcal{L}$]
 \label{thm:monomials_decay} 
The exponential asymptotic spatial decay rate $\theta$ defined in \eqref{eq:theta_def} of the Kalman gain $\mathcal{L}$ for the spatially invariant plant \eqref{eq:plant_monomial} subject to noise processes satisfying Assumption \ref{ass:A2} 
is
\begin{equation}
\theta =  \sin\bigg( \frac{\pi}{4n}\bigg) \bigg( \frac{\sigma_w}{|a| \, \sigma_v}\bigg)^{\frac{1}{2n}}.
\label{eq:monomials_ratio}
\end{equation}
\end{thm}
\begin{proof}
The Fourier symbol of $\mathcal{L}$ for the plant \eqref{eq:plant_monomial} subject to scaled white spatiotemporal noises is
$
\hat{\mathcal{L}}_{\lambda} = - |a| \lambda^{2n} + \sqrt{|a|^2 \lambda^{4n} + \frac{\sigma_w^2}{\sigma_v^2}}.
\label{eq:low_pass_L}
$
We use the Paley-Wiener Theorem \ref{thm:Paley-Wiener} to characterize the spatial decay rate of the Kalman gain.  Define the extension $\hat{\mathcal{L}}_z$ of $\hat{\mathcal{L}}_{\lambda}$ to the complex plane (Def. \ref{defn:extension}): 
\begin{equation}
    \hat{\mathcal{L}}_z \coloneqq - |a| \,  \mathbf{i}^{2n} \, z^{2n} + \sqrt{|a|^2 z^{4n} + \frac{\sigma_w^2}{\sigma_v^2}}, \; z \in \mathbb{C}.
    \label{eq:analytic_continuation_even}
\end{equation}
To determine the asymptotic spatial decay rate of $\mathcal{L}$ we find the strip in the complex plane in which $\hat{\mathcal{L}}_{z}$  is analytic.  The first term in \eqref{eq:analytic_continuation_even} is a polynomial,  an entire function.
 The second term contains a square root,  a multivalued  function in the complex plane; 
its branch points and branch cuts determine the region of the complex plane in which \eqref{eq:analytic_continuation_even} is analytic.  The branch points of \eqref{eq:analytic_continuation_even} are $|z_{\infty}| = \infty$ and the roots $z_k$ of the radicand:
\begin{equation}
    z_k = \bigg( \frac{\sigma_w}{|a| \, \sigma_v}\bigg)^{\frac{1}{2n}}  e^{\frac{(2k+1)\pi }{4n} \mathbf{i}}\,, \;  k = 0, 1, \dots, 4n-1.
     \label{eq:BPs_monomial_even}
\end{equation}
The finite branch points \eqref{eq:BPs_monomial_even} lie on a circumference centered at the origin of the complex plane,  of radius $(\sigma_w/|a| \sigma_v)^{1/2n}$: modifying plant's parameter values affects the modulus of the branch points, but their argument is preserved. The asymptotic spatial decay rate $\theta$ of the gain $\mathcal{L}$ is determined by the smallest real part (in absolute value) of the branch points \eqref{eq:BPs_monomial_even}: 
$
   \theta =\cos\Big(\frac{(2n-1) \pi}{4n}\Big) \Big( \frac{\sigma_w}{|a| \, \sigma_v}\Big)^{\frac{1}{2n}} = \sin\Big( \frac{\pi}{4n}\Big) \Big( \frac{\sigma_w}{|a| \, \sigma_v}\Big)^{\frac{1}{2n}}.
    \label{eq:decay_rate_monomial_even}
$
\end{proof}
Next, we interpret the ratio $(\sigma_w/|a| \, \sigma_v)^{1/2n}$ in \eqref{eq:monomials_ratio} and provide intuition about the spatial behavior of the feedback operator $\mathcal{L}$ accordingly.


\subsubsection{The Information Lengthscale}
\label{subsubsec:Information_lengthscale}

Characteristic lengths are valuable to set the scale of a physical system; they are useful to define dimensionless groups and  to predict properties of a dynamical system.
Dimensional analysis reveals that the parameter ratio in \eqref{eq:monomials_ratio} has units of one over length.  Hence,  for the setting of Theorem \ref{thm:monomials_decay} we define the \textit{characteristic lengthscale} or \textit{information lengthscale} of the Kalman filter as
\begin{equation}
l_* \coloneqq \bigg(\frac{2 \, |a|\, \sigma_v }{\sigma_w}\bigg)^{\frac{1}{2n}}.
\label{eq:monomial_characteristic_lengthscale}
\end{equation}
The information lengthscale $l_*$ serves as a proxy for the spatial spread of the feedback kernel $L$, that is,  as a lengthscale of useful sensory inputs for the filter.  Given a spatial location $x$,  
when $l_*$ is small,  $L$ decays fast in space and the filter weights more heavily measurements from the neighborhood of $x$ than those from sensors far away; as $l_*$ grows,  the spatial decay of $L$ slows down and measurements from farther away become increasingly relevant.  For the case of diffusive dynamics subject to spatiotemporal white process and measurement noise,  the kernel $L$ of the Kalman gain is plotted for different values of the information lengthscale $l_*$ in Fig. \ref{fig:diffusion_whiteNoises}c,  illustrating that the higher $l_*$,  the more widespread $L$ is in space.


\subsubsection{Spatial Localization and the Uncertainty Principle}
\label{subsubsec:SNR_uncertainty}

The \textit{normalized variance} of an element $f \in L_{\mathbb{C}}^2(\mathbb{R})$ is defined as $
    \mathcal{V}(f) \coloneqq \int_{\mathbb{R}} |x-m_f|^2 \, |f(x)|^2 \mathrm{d}x/\|f\|^2,
    \label{eq:normalized_variance}
$
where $
m_f \coloneqq  \int_{\mathbb{R}} x \,  |f(x)|^2 \mathrm{d}x /\|f\|^2
\label{eq:center_of_function}
$
is the center of $f$.
For a Fourier transform pair $f \in L_{\mathbb{R}}^2(\mathbb{R})$ and $\hat{f}\in L^2_{\mathbb{C}}(\mathbb{R})$  satisfying $\int_{\mathbb{R}} |x| \, f(x)^2 \mathrm{d}x< \infty$ and $\int_{\mathbb{R}} |\lambda| \, |\hat{f}(\lambda)|^2 \mathrm{d}\lambda< \infty$, \textit{Heisenberg's inequality} provides the following lower bound on the product of the respective normalized variances: 
$
    \mathcal{V}(f) \cdot \mathcal{V}(\hat{f}) \geq 1/4. 
    \label{eq:Heisenberg_inequality}
$
Qualitatively,  the inequality states that ``a non-zero function and its Fourier transform cannot be simultaneously sharply localized" \cite{Folland1997}.

In our problem, $\mathcal{V}(\hat{\mathcal{L}}_{\lambda})$ might be used as a proxy for the spread of the Fourier symbol $\hat{\mathcal{L}}_{\lambda}$ in spatial frequency. Dimensional analysis provides the scaling 
$
\mathcal{V}(\hat{\mathcal{L}}_{\lambda}) \sim (1/l_*)^2
\label{eq:monomial_varL_scaling}
$.
Then,  Heisenberg's inequality yields
$
\mathcal{V}(\mathcal{L}) \geq \frac{1}{4} \big(\mathcal{V}(\hat{\mathcal{L}}_{\lambda})\big)^{-1}.
\label{eq:Heisenberg_monomial}
$
It follows that
$
\mathcal{V}(\mathcal{L})  \sim l_*^2,
$
which is consistent with our previous interpretation of $l_*$ being the filter's information lengthscale and indicative of the spatial spread of the Kalman gain.  








\subsection{Performance} 
We aim to identify parameters that  simultaneously increase the spatial localization of the filter and improve its performance,  providing guidelines for \textit{plant and filter co-design}.

\begin{prop}[Performance of the Kalman filter]
\label{prop:monomials_performance}
Consider the plant \eqref{eq:plant_monomial} subject to noise processes satisfying Assumption \ref{ass:A2}.  
Then,  the scaling of the steady-state variance of the optimal estimation error, $\mathrm{var}(e)$, in problem parameters $(\sigma_w, \sigma_v, a)$ is
\begin{equation}
 \mathrm{var}(e)
\sim \sigma_w^{\frac{2n+1}{2n}} \sigma_v^{\frac{2n-1}{2n}}|a|^{-\frac{1}{2n}}.
\label{eq:variance_monomial_scaling}
\end{equation}
\end{prop}
\noindent \textit{Proof.}
    The scaling in \eqref{eq:variance_monomial_scaling} is simply obtained through dimensional analysis.  From \eqref{eq:Fourier_symbol_covariance},  the power spectral density of the optimal estimation error is 
\begin{align}
\hat{\mathcal{P}}_{\lambda} & = \frac{\sigma_w^2 \sigma_v^2}{|a| \sigma_v^2  \lambda^{2n} + \sqrt{|a|^2 \sigma_v^4  \lambda^{4n}  + \sigma_w^2 \sigma_v^2}} \nonumber \\
& = \frac{2 \sigma_w \sigma_v}{ \Lambda^{2n}  + \sqrt{ \Lambda^{4n}  + 4}}   = \sigma_w \sigma_v \, \mathsf{f}_n(\Lambda), 
\label{eq:PSD_monomial}
\end{align}
where we defined the \textit{dimensionless} spatial frequency $\Lambda \coloneqq \lambda \, l_*$ with $l_*$ as in \eqref{eq:monomial_characteristic_lengthscale} and the \textit{dimensionless} function
$
\mathsf{f}_n(\Lambda) \coloneqq 2 \big(\Lambda^{2n}  + \sqrt{ \Lambda^{4n}  + 4}\big)^{-1}.
$
The steady-state variance of the optimal estimation error is
\begin{align}
\mathrm{var}(e) & =  
 \frac{1}{2 \pi} \int_{-\infty}^{\infty} \hat{\mathcal{P}}_{\lambda} \, \mathrm{d}\lambda 
 \stackrel{\eqref{eq:PSD_monomial}}{=}  \frac{1}{2 \pi} \int_{\mathbb{R}} \sigma_w \sigma_v  \mathsf{f}_n(\Lambda)  \mathrm{d} \Big( \frac{\Lambda}{l_*} \Big)\nonumber \\
& = \frac{1}{2^{\frac{1}{2n}} \pi} \, \sigma_w^{\frac{2n+1}{2n}} \sigma_v^{\frac{2n-1}{2n}}|a|^{-\frac{1}{2n}} \underbrace{\int_{\mathbb{R}} \mathsf{f}_n(\Lambda) \mathrm{d}\Lambda.}_{\text{dimensionless constant}}  \; \hfill \blacksquare \nonumber
\end{align}


\eqref{eq:variance_monomial_scaling} implies that 
an improvement of sensing quality
(i.e.,  reduction of $\sigma_v$) \textit{concurrently} yields better filtering performance \eqref{eq:variance_monomial_scaling} and a more localized Kalman gain (i.e.,  a shorter  information lengthscale $l_*$ as defined in \eqref{eq:monomial_characteristic_lengthscale}). 




\section{Case Studies}
\label{sec:case_studies}
We present two case studies to illustrate our previous theoretical results.  We analyze the information structures of Kalman filters for two common spatiotemporal processes: i) a diffusion process,  and ii) a linearized Swift-Hohenberg equation,  both over the real line.  Through these examples we introduce the \textit{branch point locus (BPL)},  a useful tool to analyze the sensitivity of the spatial localization of the Kalman gain to system parameters.   

\subsection{Diffusion on the real line}
\label{subsec:diffusion}

We study the Kalman filter for a diffusion process in the real line with spatially distributed measurements: 
\begin{subequations}
\begin{align}
\frac{d \boldsymbol{\psi}}{d t} (t) & = \mathcal{A} \boldsymbol{\psi}(t) + \sigma_w \boldsymbol{w}(t), \label{eq:diffusion_dynamics}\\
\boldsymbol{y}(t) & = \boldsymbol{\psi}(t) + \mathcal{G} \boldsymbol{v}(t), \label{eq:diffusion_measurement}
\end{align}
\label{eq:diffusion_plant}
\end{subequations}
where $\mathcal{A}:= \kappa \, \partial_x^2$
\text{ with }
$\mathcal{D}(\mathcal{A}) = H_2(\mathbb{R})$ and 
$\kappa>0$ the diffusivity constant. $\sigma_w>0$ and the noise processes satisfy Assumption \ref{ass:A2}. 
The plant \eqref{eq:diffusion_plant} fits the abstract state-space representation \eqref{eq:plant_abstract} with $\mathcal{C} = \mathcal{I}$.  The operator $\mathcal{A}$  
 is 
spatially invariant,  with Fourier symbol $\hat{\mathcal{A}}_{\lambda} = - \kappa \lambda^2$.  Hence,  $\mathcal{A}$ is the infinitesimal generator of a $C_0$-semigroup in $L^2(\mathbb{R})$ by Proposition \ref{prop:C0_generation}.  $\mathcal{G}$ is a spatially invariant operator on $L^2(\mathbb{R})$ and we consider two settings: in the first,  $\mathcal{G}$ scales the spatiotemporally white measurement noise; in the second,  $\mathcal{G}$ introduces short-range spatial autocorrelations in the measurement noise.  The plant \eqref{eq:diffusion_plant} fits the framework of Section \ref{sec:spatial_llocalization_and_performance} and will serve as an illustrative example.  We study the Kalman filter for \eqref{eq:diffusion_plant} in the spatial frequency domain.
This case study extends the results presented in \cite{Arbelaiz:2020}.\\
\vspace{-0.25cm}

\subsubsection{Scaled spatiotemporal white noise processes}
\label{subsubsec:diffusion_white_noises}
Let $\mathcal{G} = \sigma_v\, \mathcal{I}$ in \eqref{eq:diffusion_plant},  with $\sigma_v >0$.  
In this setting, the Kalman gain has Fourier symbol 
$
\hat{\mathcal{L}}_{\lambda} = -\kappa  \lambda^2 + \sqrt{\kappa^2 \lambda^4 + \frac{\sigma_w^2}{\sigma_v^2}} 
\label{eq:diffusion_KalmanGain_FourierSymbol_whiteNoises}
$
with extension
\begin{equation}
\hat{\mathcal{L}}_z \coloneqq \kappa \, z^2 + \sqrt{\kappa^2 z^4 + \frac{\sigma_w^2}{\sigma_v^2}}, \; z \in \mathbb{C}.
\label{eq:diffusion_KalmanGain_FourierSymbolExtension_whiteNoises}
\end{equation}
Following Section \ref{subsubsec:Information_lengthscale},  the \textit{information lengthscale} is
\begin{equation}
l_* \coloneqq  \bigg( \frac{2 \, \kappa \, \sigma_v}{\sigma_w} \bigg)^{\frac{1}{2}}.
\label{eq:diffusion_information_lengthscale}
\end{equation}

\begin{figure}[h]
\centering
\includegraphics[width=\linewidth]{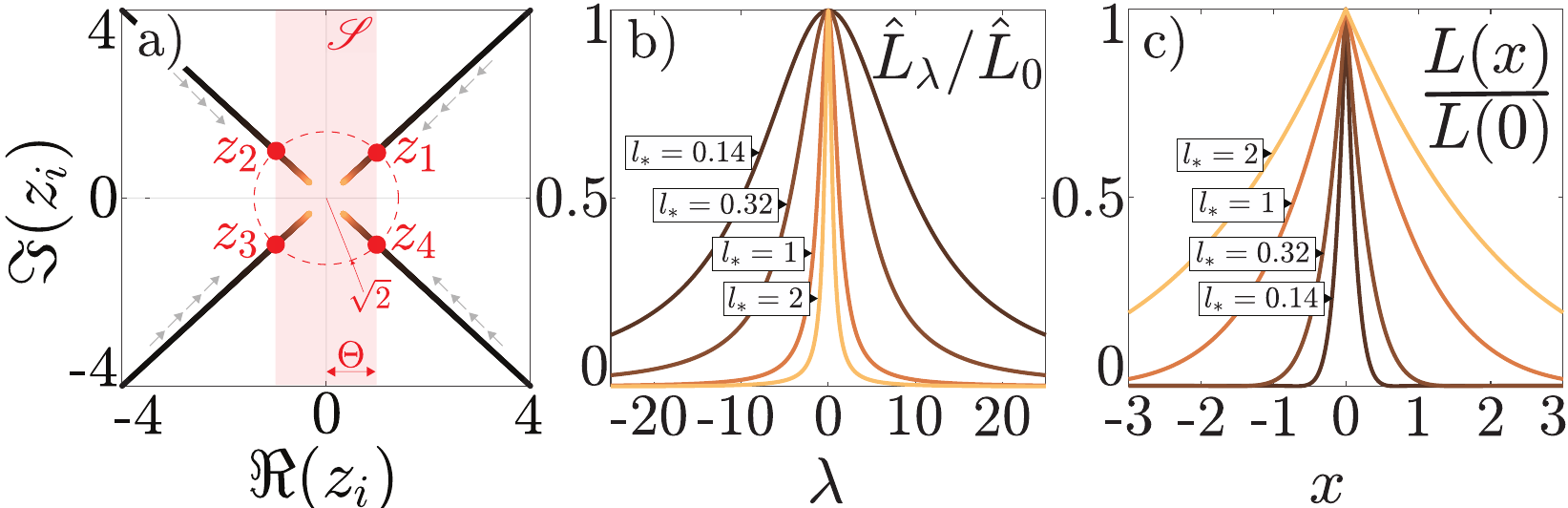}
\caption{ \small{Diffusion with scaled spatiotemporally white process and measurement noises.  Color-code is consistent among panels.  a) BPL: trajectories of the branch points \eqref{eq:diffusion_BP_whiteNoises} as $l_*$ is varied (branch cuts omitted).  Arrows indicate the direction of increasing $l_*$.  Branch points $z_{1,2,3,4}$ and analyticity strip $\mathscr{S}$ for $l_*= 1$ in red.  b) Normalized Fourier transform $\hat{L}_{\lambda}/\hat{L}_0$  against spatial frequency $\lambda$ for different values of $l_*$,  as indicated.   c) Normalized Kalman gain kernel $L(x)/L(0)$ against the spatial coordinate $x$, for different values of $l_*$, as given.  
}}
\label{fig:diffusion_whiteNoises}
\end{figure}

\paragraph*{Spatial Localization of the Kalman gain $\mathcal{L}$} 
The branch points of \eqref{eq:diffusion_KalmanGain_FourierSymbolExtension_whiteNoises} are $|z_{\infty}| = \infty$ and 
\begin{equation}
z_n = \bigg( \frac{\sigma_w}{\kappa \, \sigma_v}\bigg)^{\frac{1}{2}}e^{\frac{(2n-1) \pi}{4}\mathbf{i}},  \text{ with } n=1,2,3,4.
\label{eq:diffusion_BP_whiteNoises}
\end{equation}
More compactly,  $z_{1,2,3,4} = l_*^{-1} \big( \pm 1 \pm i\big)$. The asymptotic decay rate $\theta$ of the Kalman gain $\mathcal{L}$ is determined by the real part of \eqref{eq:diffusion_BP_whiteNoises},  that is, $\theta = l_*^{-1}$, which is consistent with Theorem \ref{thm:monomials_decay}: the higher the information lengthscale $l_*$ is,  the more widespread $\mathcal{L}$ is in space,  as shown in Fig. \ref{fig:diffusion_whiteNoises}c).  

\paragraph*{Filtering Performance}  The power spectral density of the optimal estimation error is 
\begin{equation}
\hat{\mathcal{P}}_{\lambda} =  \frac{\sigma_w^2 /\kappa}{\lambda^2 + \sqrt{\lambda^4 + \frac{\sigma_w^2}{\kappa^2 \sigma_v^2}}},
\label{eq:PSD_diffusion_white_noise}
\end{equation}
and consequently, its steady-state variance $\mathrm{var}(e)$   is
\begin{equation}
\mathrm{var}(e) = \frac{1}{2\pi} \int_{-\infty}^{\infty} \hat{\mathcal{P}}_{\lambda} \mathrm{d}\lambda \,  \stackrel{\eqref{eq:PSD_diffusion_white_noise}}{=} \, \frac{1}{6 \pi^{\frac{3}{2}}} \frac{\sigma_w^{\frac{3}{2}} \sigma_v^{\frac{1}{2}}}{\kappa^{\frac{1}{2}}} \, \Gamma\bigg(\frac{1}{4}\bigg)^2,
\label{eq:diffusion_uncorrelated_cost}
\end{equation}
where $\Gamma(\boldsymbol{\cdot})$ denotes the Gamma function.  The scaling in problem parameters obtained in \eqref{eq:diffusion_uncorrelated_cost} through integration is consistent with that provided by Proposition \ref{prop:monomials_performance} (for $n=1$),  which was simply derived using dimensional analysis.  \\
\vspace{-0.25cm}

\paragraph*{The Branch Point Locus (BPL)} 
The root locus is a useful tool to analyze the stability of a  linear system through visual inspection of the trajectories of its poles on the complex temporal $s$-plane as a parameter is varied.  In an analogous manner,  we define the \textit{branch point locus (BPL)} as the trajectories of the branch points of the extension $\hat{\mathcal{L}}_z$ in the complex spatial $z$-plane as a parameter of interest is changed.  The BPL facilitates to visually identify the values of the parameter yielding a highly spatially localized Kalman filter.  Fig. \ref{fig:diffusion_whiteNoises}a) illustrates the BPL  for the plant \eqref{eq:diffusion_plant} subject to white noises as $l_*$ is varied.  It also shows the analyticity strip $\mathscr{S}$ of \eqref{eq:diffusion_KalmanGain_FourierSymbolExtension_whiteNoises} for $l_* = 1$,  that is,  $\mathscr{S} = \{ z \in \mathbb{C}: |\Re(z)|< 1 \}$.
 We remark the usefulness of dimensional analysis to define the unique parameter $l_*$,  which allows us to plot the BPL:  visualization of the branch point trajectories is cumbersome if instead the effect of variations in each parameter of the plant is independently considered. 
 Fig. \ref{fig:diffusion_whiteNoises}b)-c) exhibit the Fourier transform pair of the normalized Kalman gain for the plant \eqref{eq:diffusion_plant},
showcasing  the \textit{uncertainty principle} described in Section \ref{subsubsec:SNR_uncertainty}: when the gain is localized in the physical domain,  it becomes widespread in frequency,  and viceversa.  \\
\vspace{-0.25cm}

\subsubsection{Spatially autocorrelated measurement noise} 
We consider the setting in which $\mathcal{G}$ introduces \textit{short-range autocorrelations in space,} that is,
$|\hat{\mathcal{G}}_{\lambda} |^2
=\frac{\sigma_v^2}{1+l_v^2 \lambda^2},$ where $l_v$ denotes the \textit{autocorrelation length} of the measurement noise. 

 
\paragraph*{Dimensional Analysis}  Dimensional analysis allows us to group the parameters of the plant in a reduced number of \textit{dimensionless parameters} with physical significance.  By Buckingham's $\Pi$-theorem,  there is a \textit{unique} dimensionless parameter $\Pi_*$ in our set-up,  that we define as
\begin{equation}
\Pi_* \coloneqq \frac{l_v}{l_*},
\label{eq:diffusion_dimensionless_number}
\end{equation}
where $l_*$ is as in \eqref{eq:diffusion_information_lengthscale}.  $\Pi_*$ is the ratio of the two lengthscales in the problem.  Any dimensionless quantity can be written as a function of $\Pi_*$.   We denote $\Pi_* = 1$ by $\Pi_*^m$ and refer to it as a \textit{matching condition},  since the values of the two lengthscales ``match".   
We define the dimensionless 
variables  
$
\Psi \coloneqq   \frac{\psi}{\sigma_v},  \;
 \mathsf{t} \coloneqq \frac{t \, \sigma_w}{2 \, \sigma_v},  \; 
 \mathsf{x} \coloneqq \frac{x}{l_*},  \; 
 \mathsf{v}  \coloneqq v,  \; 
 \mathsf{w} \coloneqq 2\, w.
$
Accordingly, the dimensionless counterpart of the plant \eqref{eq:diffusion_plant} is
\begin{subequations}
\begin{align}
\frac{\partial \Psi}{ \partial \mathsf{t}}(\mathsf{x}, \mathsf{t}) & = \frac{\partial^2 \Psi}{\partial \mathsf{x^2}} (\mathsf{x}, \mathsf{t}) +  \mathsf{w}(\mathsf{x}, \mathsf{t}),  \label{eq:diffusion_dynamics_dimensionless}\\
\mathsf{y}(\mathsf{x}, \mathsf{t}) & = \Psi (\mathsf{x}, \mathsf{t}) + \mathsf{v}(\mathsf{x}, \mathsf{t}). \label{eq:diffusion_measurement_dimensionless}
\end{align}
\label{eq:diffusion_plant_dimensionless}
\end{subequations}
\noindent The dimensionless 
symbol $\hat{\mathsf{L}}_{\Lambda}$ of the  Kalman gain is
\begin{equation}
\hat{\mathsf{L}}_{\Lambda} = - \Lambda^2 + \sqrt{\Lambda^4 + 4 \Pi_*^2 \Lambda^2 + 4},
\label{eq:diffusion_KalmanGain_FourierSymbol_dimensionless}
\end{equation}
where $ \Lambda \coloneqq \lambda \,  l_*,  $ is the dimensionless spatial frequency.  Had we non-dimensionalized the plant in the example of Section \ref{subsubsec:diffusion_white_noises},  the dimensionless decay rate $\Theta$ of the Kalman gain would have been $\Theta = 1$; thus, the kernels illustrated in Fig.  \ref{fig:diffusion_whiteNoises} would have collapsed into a single one when plotted against the corresponding dimensionless variable (namely,  $\Lambda$ or $\mathsf{x}$). 

\begin{figure}[h]
\includegraphics[width=\linewidth]{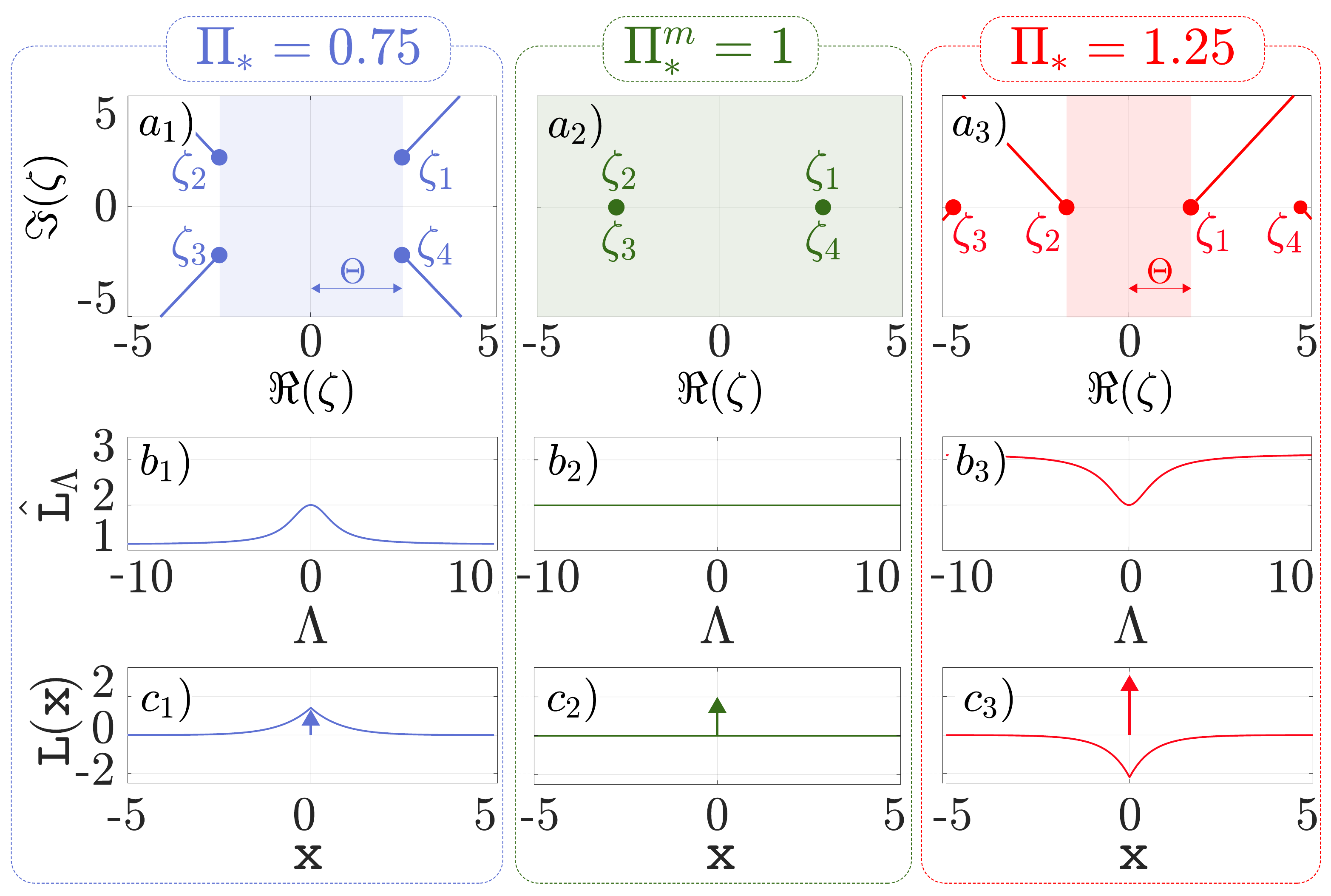}
\caption{\small{Diffusion with spatially autocorrelated measurement noise.  Colors according to different values of $\Pi_*$, as indicated.  a) Branch points,  branch cuts,  and analyticity region of the extension of \eqref{eq:diffusion_KalmanGain_FourierSymbol_dimensionless} in the complex plane.  Kalman gain operator: b)  $\hat{\mathsf{L}}_{\Lambda}$ as a function of spatial frequency $\Lambda$,  and c) $\mathsf{L}$ as a function of the spatial coordinate $\mathsf{x}$.  All variables are dimensionless.  Vertical axis is preserved in each row.  
}}
\label{fig:diffusion_correlated}
\end{figure}

\begin{figure}[h]
\includegraphics[width=\linewidth]{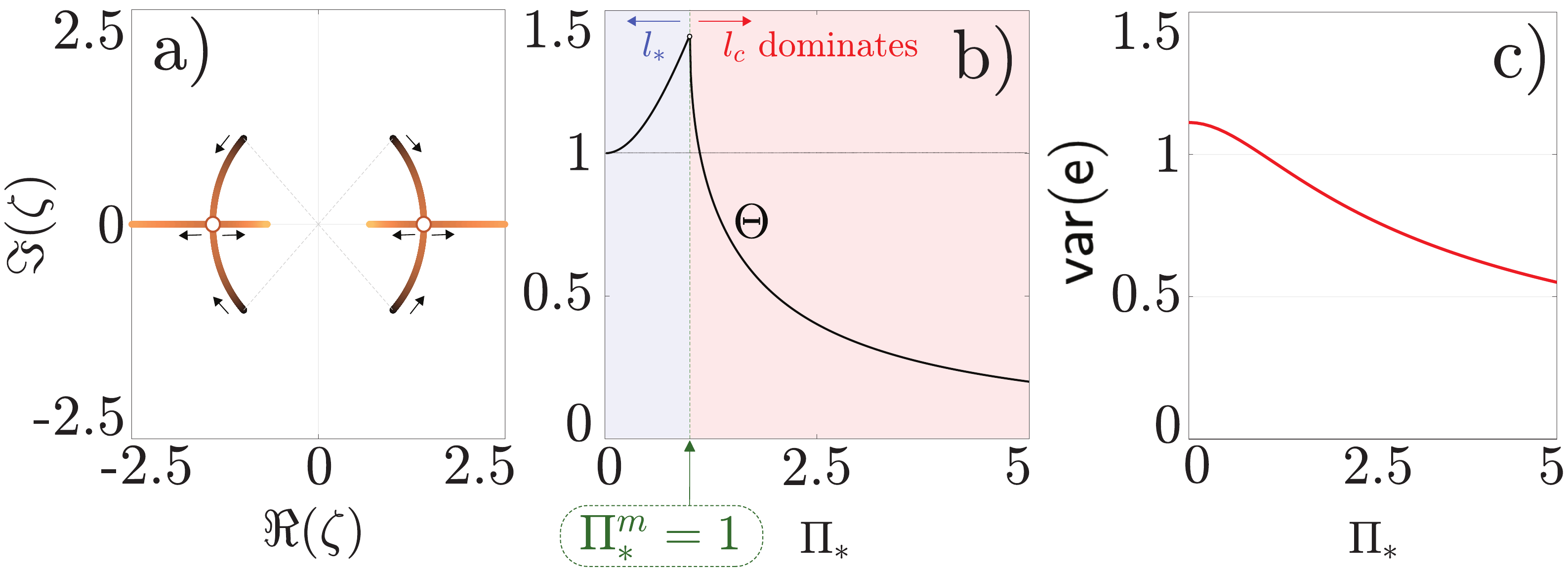}
\caption{\small{Diffusion with spatially autocorrelated measurement noise.  a) BPL of the extension of \eqref{eq:diffusion_KalmanGain_FourierSymbol_dimensionless}, colored according to $\Pi_*$.  Arrows indicate increasing $\Pi_*$.  b) Dimensionless decay rate $\Theta$ \eqref{eq:diffusion_dimensionless_decayRate} of $\mathcal{L}$.  The area shaded in blue (red) corresponds to $\Theta$ being dominated by $l_*$ ($l_v$).  c) Steady-state dimensionless performance $\mathsf{var}(\mathsf{e})$ of the Kalman filter  plotted as a function of $\Pi_*$.  
}}
\label{fig:diffusion_correlated_cost}
\end{figure}

\paragraph*{Spatial Localization of the Kalman gain $\mathcal{L}$} In the current problem set-up,   the Kalman gain consists of (see Fig. \ref{fig:diffusion_correlated}c)
\begin{itemize}
\item a \textit{Dirac delta distribution} located at the origin; and
\item a component with \textit{asymptotic exponential spatial decay}.
\end{itemize} 
The strength of the Dirac distribution 
is determined by the value of the horizontal asymptote of  $\hat{\mathsf{L}}_{\Lambda}$:
\begin{equation}
\lim_{|\Lambda| \to \infty} \hat{\mathsf{L}}_{\Lambda} = \lim_{|\Lambda| \to \infty}  \frac{4\big( \Pi_*^2 \Lambda^2 + 1\big)}{\Lambda^2 + \sqrt{\Lambda^4 + 4 \Pi_*^2 \Lambda^2 + 4}} = 2 \, \Pi_*^2.
\label{eq:diffusion_correlated_delta_strength}
\end{equation}
To determine the asymptotic decay rate of the exponentially decaying component of the kernel as a function of $\Pi_*$,  we proceed analogously to the previous sections and define the extension of \eqref{eq:diffusion_KalmanGain_FourierSymbol_dimensionless} to the complex plane.  Its dimensionless complex finite branch points $\zeta_i \in \mathbb{C}$ are
\begin{equation}
\zeta_{1,2,3,4}(\Pi_*) = 
\begin{cases}
\pm \sqrt{1 + \Pi_*^2} \pm \mathbf{i} \sqrt{1 - \Pi_*^2}, \;  \text{ if } 0 \leq \, \Pi_* < 1,\\
\pm \sqrt{2 \big(\Pi_*^2 \pm \sqrt{\Pi_*^4 - 1} \big)}, \,\,\; \;\text{ if } \Pi_* > 1.\\
\end{cases}
\label{eq:diffusion_BPs_correlated}
\end{equation}
Accordingly,  the dimensionless asymptotic spatial decay rate $\Theta$ of $\mathcal{L}$ is
\begin{equation}
\Theta = 
\begin{cases}
\sqrt{1 + \Pi_*^2}, \;  \hspace{1.6cm} \text{ if } 0 \leq \Pi_* < 1, \\
\sqrt{2 \big(\Pi_*^2 - \sqrt{\Pi_*^4 - 1} \big)}, \text{ if } \Pi_* >1.
\end{cases}
\label{eq:diffusion_dimensionless_decayRate}
\end{equation}
Some interesting observations related to the decay rate $\Theta$ are:
\textit{(1)} $\Theta$ is \textit{not} monotonic in $\Pi_*$ (see Fig.  \ref{fig:diffusion_correlated_cost}b).   For $\Pi_* < \Pi_*^m$,  $l_*$ is the \textit{dominating} lengthscale: given a location $\mathsf{x}$, $l_*$ dictates how far away to get measurements from for the filtering task at $\mathsf{x}$.    When $\Pi_* > \Pi_*^m$,  $l_v$ dominates and sets the useful information lengthscale for the filter. 
The matching value $\Pi_*^m = 1$  is \textit{critical} in the sense that the Kalman filter is \textit{completely decentralized} (see Def. \ref{defn:completely_decentralized} and Fig. \ref{fig:diffusion_correlated}).  At $\Pi_*^m$ the branch points \eqref{eq:diffusion_dimensionless_decayRate} transition from complex conjugates to real,  collapsing pairwise in the real axis.  The extension $\hat{\mathsf{L}}_{\zeta}(\Pi_*^m)$ is entire.  Indeed,  $\hat{\mathsf{L}}_{\zeta}(\Pi_*^m) = 2$.   A similar spatial behavior was reported in Kalman filters for elastic wave dynamics \cite{Arbelaiz2022};
\textit{(2)} The sign of the exponentially decaying component of the feedback kernel flips at $\Pi_*^m$: for $\Pi_*< \Pi_*^m$,  neighboring measurements are fed back with a positive sign; for $\Pi_*< \Pi_*^m$, with a negative sign (see Fig.  \ref{fig:diffusion_correlated}c);
\textit{(3)} The Kalman gain can be more widespread in space when measurement noise is spatially autocorrelated (see Fig. \ref{fig:diffusion_correlated_cost}b); hence, it is important to take autocorrelations into account (which are often disregarded in the literature) when designing decentralized filter architectures.

\paragraph*{Filtering performance} The dimensionless power spectral density of the optimal estimation error is
\begin{equation}
\hat{\mathsf{P}}_{\Lambda}( \Pi_*) = \frac{4}{\Lambda^2 + \sqrt{\Lambda^4 + 4 \Pi_*^2 \Lambda^2 + 4}}.
\end{equation}
Then,  $\mathrm{d} \hat{\mathsf{P}}_{\Lambda}/\mathrm{d} \Pi_* \leq 0,  \; \forall \Lambda$.  Since $\mathsf{var}(\mathsf{e}; \Pi_*) =  \frac{1}{2 \pi}   \int_{\mathbb{R}} \hat{\mathsf{P}}_{\Lambda} \, \mathrm{d} \Lambda$,  by Leibniz integral rule 
$\mathsf{var}(\mathsf{e}; \Pi_*)$ monotonically decreases with $\Pi_*$ (see Fig. \ref{fig:diffusion_correlated_cost}c). 



\vspace{-0.3cm}
\subsection{Linearized Swift-Hohenberg equation on the real line}
\label{subsec:example_2}
The Swift-Hohenberg equation is a normal form for distributed systems 
that exhibit spatially localized structures.  We study the Kalman filter for its linearization:  
\begin{subequations}
\begin{align}
\frac{d \boldsymbol{\psi}}{d t}(t) & = \mathcal{A} \boldsymbol{\psi}(t) + \sigma_w \boldsymbol{w}(t), \label{eq:dynamics_Swift_Honenberg}\\
\boldsymbol{y}(t) & = \boldsymbol{\psi}(t) + \sigma_v  \boldsymbol{v}(t),  \label{eq:measurement_Swift_Honenberg}
\end{align}
\label{eq:plant_Swift_Henenberg}
\end{subequations}
where the operator  
$
\mathcal{A} \coloneqq -\big(\partial_x^2
 + \frac{1}{l_A^2}\big)^2   
$
has domain 
$
\; \mathcal{D}(\mathcal{A}) = H_4(\mathbb{R})
$, and 
$l_A$ sets a preferential lengthscale of the dynamics. The noise processes $\boldsymbol{w}$ and $\boldsymbol{v}$ satisfy Assumption \ref{ass:A2}.  
The operator $\mathcal{A}$ is spatially invariant,  with Fourier symbol 
$\hat{\mathcal{A}}_{\lambda} = - \big(\lambda^2 - \frac{1}{l_A^2}\big)^2
\label{eq:Swift_Hohenberg_FourierSymbol}
$.  By Proposition \ref{prop:C0_generation},  
$\mathcal{A}$ is the infinitesimal generator of a $C_0$-semigroup in $L^2(\mathbb{R})$.  The plant \eqref{eq:plant_Swift_Henenberg} does not fit the framework of Section \ref{sec:spatial_llocalization_and_performance},  as $\hat{\mathcal{A}}_{\lambda}$ is not an even power of a monomial; however,  it follows the more general problem set-up of Section \ref{sec:Problem_Formulation} and hence,  we use Theorem \ref{thm:KB_freq} to study the Kalman filter.  \\
\vspace{-0.25cm}
\paragraph*{Non-dimensionalization} Define the lengthscale\footnote{The operator $\mathcal{A}$ in the plant \eqref{eq:plant_Swift_Henenberg}
is $\mathcal{A} =-a \big(\partial_x^2
 + 1/l_A^2\big)^2$ with $a = 1$,  where $a$ has units of length$^4$ over time.  These units must be taken into account for dimensional consistency in the definition of the lengthscale \eqref{eq:Swift_Honenberg_information_lengthscale}.}
\begin{equation}
l_* \coloneqq  \bigg( \frac{2 \, \sigma_v}{\sigma_w}\bigg)^{\frac{1}{4}},
\label{eq:Swift_Honenberg_information_lengthscale}
\end{equation}
 and the dimensionless variables $\Psi = \psi/\sigma_v$,  $\mathsf{t} \coloneqq t \, \sigma_w/2 \sigma_v$,  $\mathsf{x} \coloneqq x/l_*$, $\mathsf{w} \coloneqq 2\, w $, $\mathsf{v} \coloneqq v$.  By Buckingham's $\Pi$-theorem there is a \textit{unique} dimensionless group $\Pi_*$ in this setting. We define it as the ratio of the two lengthscales of the problem 
\begin{equation}
\Pi_* \coloneqq \frac{l_A}{l_*}.
\label{eq:pi_swift_hohenberg}
\end{equation}
The dimensionless counterpart of the plant \eqref{eq:plant_Swift_Henenberg} is
\begin{subequations}
\begin{align}
\frac{\partial \Psi}{\partial \mathsf{t}} (\mathsf{x}, \mathsf{t}) & = -\bigg( \partial_{\mathsf{x}}^2 + \frac{1}{\Pi_*^2}\bigg)^2 \Psi  (\mathsf{x}, \mathsf{t}) + \mathsf{w}  (\mathsf{x}, \mathsf{t}),\label{eq:dynamics_Swift_Hohenberg_dimensionless}\\
\mathsf{y} (\mathsf{x}, \mathsf{t}) & = \Psi  (\mathsf{x}, \mathsf{t}) + \mathsf{v} (\mathsf{x}, \mathsf{t}). \label{eq:measurement_Swift_Hohenberg_dimensionless}
\end{align}
\label{eq:plant_Swift_Hohenberg_dimensionless}
\end{subequations}
The dimensionless symbol of the Kalman gain for  \eqref{eq:plant_Swift_Hohenberg_dimensionless} is 
\begin{equation}
\hat{\mathsf{L}}_{\Lambda} = - \bigg( \Lambda^2 - \frac{1}{\Pi_*^2}\bigg)^2 + \sqrt{\bigg(  \Lambda^2 - \frac{1}{\Pi_*^2}\bigg)^4 + 4},
\label{eq:Swift_Honenberg_whiteNoise_L_symbol}
\end{equation}
where $\Lambda  \coloneqq \lambda \,  l_*$ is the  dimensionless spatial frequency.  Since $\lim_{|\Lambda| \to \infty} \hat{\mathsf{L}}_{\Lambda} = 0$,  the gain does not contain point-supported distributions and has exponential spatial asymptotic decay.\\
\vspace{-0.25cm}

\paragraph*{Spatial localization of the Kalman gain $\mathcal{L}$}
To characterize the exponential asymptotic spatial decay rate of \eqref{eq:Swift_Honenberg_whiteNoise_L_symbol},  we define its analytic extension to a region of  the complex plane by 
\begin{equation}
\hat{\mathsf{L}}_{\zeta} \coloneqq - \bigg( \zeta^2 + \frac{1}{\Pi_*^2}\bigg)^2  + \sqrt{\bigg( \zeta^2 + \frac{1}{\Pi_*^2} \bigg)^4 + 4 },
\label{eq:Swift_Hohenberg_whiteNoise_extension}
\end{equation}
where $\zeta$ is the dimensionless complex spatial frequency.  Its branch points are $|\zeta_{\infty}| = \infty$,  and the complex conjugate pairs
$
\zeta_{1,2,3,4}  = \pm \Omega \pm (2 \Omega)^{-1} \mathbf{i} \text{ and }
\zeta_{5,6,7,8}  = \pm \Theta \pm (2 \Theta)^{-1}  \mathbf{i},
$
with
\begin{subequations}
\begin{align}
\Omega  & \coloneqq \sqrt{\frac{1}{2} \bigg[ 1 - \frac{1}{\Pi_*^2}+ \sqrt{\bigg(  1 - \frac{1}{\Pi_*^2} \bigg)^2 + 1}\bigg] }, \label{eq:Swift_Hohenberg_Omega}\\
\Theta & \coloneqq \sqrt{\frac{1}{2} \bigg[ -1 - \frac{1}{\Pi_*^2}+ \sqrt{\bigg(  1 + \frac{1}{\Pi_*^2} \bigg)^2 + 1}\bigg] }. 
\label{eq:Swift_Hohenberg_Theta}
\end{align}
\label{eq:SH_Omega_Theta}
\end{subequations}
The dimensionless asymptotic spatial decay rate of $\mathsf{L}$,  $\Theta$  monotonically increases with $\Pi_*$ (see Fig. \ref{fig:Swift_whiteNoise}c, details in Appendix 
\hyperref[sec:branch_points_Swift_white]{II}): the higher $\Pi_*$,  the higher the spatial localization of the Kalman gain.  
Furthermore,  
$
\Theta \xrightarrow[]{\Pi_* \to \infty} \sqrt{\frac{1}{2} \big( \sqrt{2}-1\big)},
$
which is consistent with the scaling reported in Theorem \ref{thm:monomials_decay} (for $n=2,  a = 1$),  with $\theta = \Theta \, l_*$ and $l_*$ as in \eqref{eq:Swift_Honenberg_information_lengthscale}  -- $\hat{\mathcal{A}}_{\Lambda}$ tends to a fourth order monomial as $\Pi_* \to \infty$,  see Fig. \ref{fig:Swift_whiteNoise}a.

\begin{figure}[h]
\centering
\includegraphics[width=0.65\linewidth]{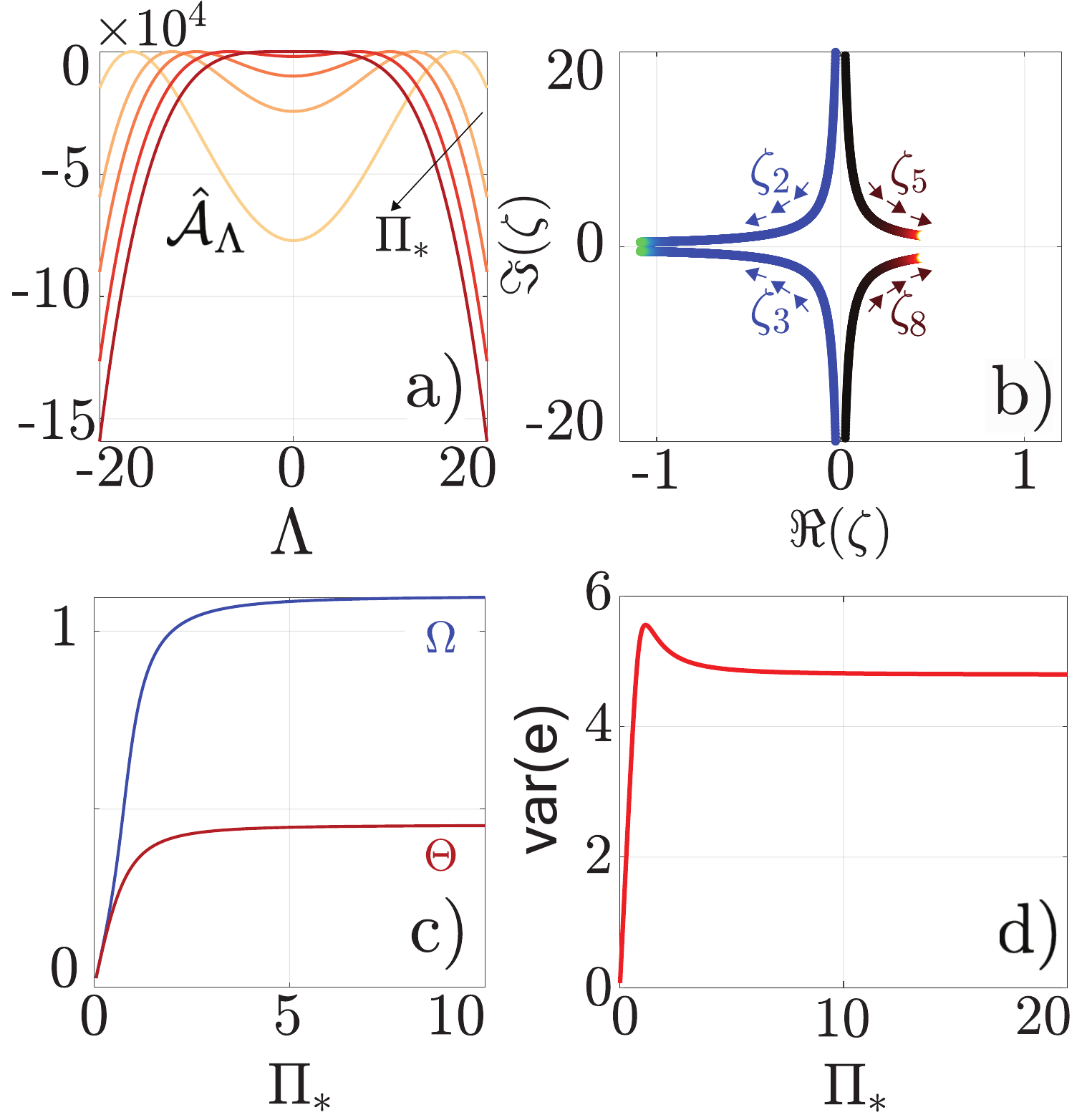}
\caption{
\small{Swift-Hohenberg dynamics with scaled white noises.   a) Fourier symbol $\hat{\mathcal{A}}_{\Lambda}$ for different values of $\Pi_*$ (darker color corresponds  to higher $\Pi_*$ value) plotted as a function of $\Lambda$.  
b) Trajectories of some of the branch points (as indicated) as a function of $\Pi_*$.  Arrows indicate direction of increasing $\Pi_*$.  c) $\Omega$  and $\Theta$ as defined in \eqref{eq:SH_Omega_Theta}.  
(d) Filter performance as a function of $\Pi_*$.
Variables in all panels are dimensionless.  
}}
\label{fig:Swift_whiteNoise}
\end{figure}




\paragraph*{Filtering performance}
The power spectral density is 
\begin{equation}
    \hat{\mathsf{P}}_{\Lambda} = \frac{4}{\big( \Lambda^2 - \frac{1}{\Pi_*^2}\big)^2 + \sqrt{\big( \Lambda^2 - \frac{1}{\Pi_*^2} \big)^4 + 4}},
\end{equation}
and  the filtering performance is $\mathsf{var}(\mathsf{e}; \Pi_*) = \frac{1}{2 \pi} \int_{\mathbb{R}} \hat{\mathsf{P}}_{\Lambda} \mathrm{d} \Lambda$, illustrated as a function of $\Pi_*$ in Fig. \ref{fig:Swift_whiteNoise}d.

\section{Conclusion \& Future Work}
\label{sec:conclusion}
\textit{Conclusion:} We analyzed the distributed-parameter Kalman filtering problem for spatiotemporal random fields governed by shift invariant dynamics over unbounded spatial domains, subject to stationary process and measurement noises.  
In this set-up, 
the 
filter is shift invariant and the Kalman gain is a spatially localized convolution.  
We explicitly characterized its degree of spatial localization as a function of the variances and autocorrelations of the noise processes perturbing the dynamics and measurements, and provided a condition under which the Kalman filter becomes completely decentralized. We argued that typically,  measurement noise must exhibit spatial autocorrelations for this condition to hold, highlighting the importance of accounting for spatial autocorrelations of the measurement noise in the filter synthesis.  
We showed the usefulness of dimensional analysis and defined the information lengthscale of the Kalman filter as a proxy for its spatial localization. We introduced a novel graphical technique, which we term the Branch Point Locus, to systematically explore the sensitivity of the spatial localization of the filter to plant's parameters.  
The new results and methods were illustrated through the analysis of two case studies: estimation of the state in  diffusion and in the linearized Swift-Hohenberg dynamics over the real line, which provided further insights. 

\textit{Future Work:} 
One of the applications that motivated this work is soft robotics. On-going research includes utilizing the fundamental insights obtained in this work to inform the design of optimal distributed control and estimation feedback architectures for soft and modular crawling robots \cite{Arbelaiz_crawling}, which use the propagation of peristaltic waves (i.e., longitudinal contractions and expansions) along their bodies for locomotion. 
On-going work also includes generalizing the characterization of the structural properties presented in this paper 
to optimal risk-aware controllers and estimators for spatially invariant systems \cite{Arbelaiz2023}, and extending the Branch Point Locus graphical tool to account for multiple spatial dimensions and time-dependent feedback operators.

\section*{Appendix I: SIS \& structural properties of Kalman filters}
\label{sec:Appendix_II}

\subsection{The half-plane condition}
\label{subsec:half_plane}

The following proposition provides a simple equivalent condition for a time independent and spatially invariant operator to be the generator of a $C_0$ semigroup in $L^2(\mathbb{R})$ in terms of its Fourier symbol.

\begin{prop}[Equivalent condition for a spatially invariant operator to be the generator of a $C_0$-semigroup in $L^2(\mathbb{R})$] \label{prop:C0_generation} 
Consider the system $\partial_t \boldsymbol{\psi}(t) = \mathcal{A} \boldsymbol{\psi}(t)$ with $\mathcal{A}: \mathcal{D}(\mathcal{A}) \subset L^2(\mathbb{R}) \to L^2(\mathbb{R})$ densely defined,  spatially invariant and time independent, with continuous Fourier symbol  $\hat{\mathcal{A}}_{\lambda}$.  Then,  
\begin{equation}
\sup_{\lambda \in \mathbb{R}} \; \Re(\hat{\mathcal{A}}_{\lambda}) \leq a_* < \infty
\label{eq:half_plane_condition}
\end{equation}
is equivalent to $\mathcal{A}$ being the generator of a $C_0$-semigroup $\mathcal{T}(t)$ in $L^2(\mathbb{R})$.  We refer to \eqref{eq:half_plane_condition} as the half-plane condition.
\end{prop}

Proposition \ref{prop:C0_generation} follows from the \textit{spectrum determined growth assumption} -- see, e.g., \cite[page 74]{Curtain:1978}, \cite[Theorem 3.1.6]{Curtain2020}.

\subsection{Structural properties of Kalman filters for SIS}
\label{subsec:structural_properties_KF}
We present important structural properties of Kalman filter for SIS needed to derive our main results.
The steady-state optimal error covariance operator $\mathcal{P}$ satisfies the 
 infinite-dimensional ARE \cite[Theorem 6.8.3]{Bala:1981}, \cite{Zabczyk1975}:
\begin{align}
& \langle \big( \mathcal{A} \mathcal{P} + \mathcal{P} \mathcal{A}^* + \mathcal{B} \mathcal{B}^* - \mathcal{P} \mathcal{C}^* (\mathcal{G} \mathcal{G}^*)^{-1} \mathcal{C} \mathcal{P}\big) h,  f \rangle = 0,  
\label{eq:estimation_OARE}
\end{align}
for all $ h, f \in \mathcal{D}(\mathcal{A}^*)$. 
In our spatially invariant setting it is more convenient to work with the Fourier transform of the ARE \eqref{eq:estimation_OARE}, in the spatial frequency domain \cite{Bamieh:2002}.  Lemma \ref{lem:spatial_invariance_P} shows that the covariance operator $\mathcal{P}$ is \textit{spatially invariant}.  
Building upon this result, Theorem \ref{thm:KB_freq}
shows that the infinite-dimensional Kalman filter can be readily synthesized in the spatial frequency domain by solving a family of finite-dimensional scalar AREs parametrized by the spatial frequency $\lambda$.


\begin{lem}[Spatial invariance of the covariance operator $\mathcal{P}$ of the optimal state estimation error] 
\label{lem:spatial_invariance_P}
Under Assumptions \ref{ass:A2} to \ref{ass:5} the steady-state covariance operator $\mathcal{P}$ of the optimal state estimation error is spatially invariant.
\end{lem}
\begin{proof}
 Under Assumptions \ref{ass:A2} to \ref{ass:5},  
the
 Algebraic Riccati Equation \eqref{eq:estimation_OARE} has a unique self-adjoint solution $\mathcal{P}$ such that $\mathcal{A} - \mathcal{P} \mathcal{C}^* (\mathcal{G} \mathcal{G}^*)^{-1} \mathcal{C}$ generates an exponentially stable semigroup \cite{Bala:1981,Zabczyk1975}.  We aim to show that in our problem set-up such a $\mathcal{P}$ is spatially invariant, that is,
$
\mathcal{T}_z \mathcal{P} f = \mathcal{P} \mathcal{T}_z f \; \Leftrightarrow \;  \mathcal{P} f = \mathcal{T}_z^{-1} \mathcal{P} \mathcal{T}_z f =  \mathcal{T}_z^* \mathcal{P} \mathcal{T}_z f,
$
for all $ f\in \mathcal{D}(\mathcal{P})$ and for all translations $\mathcal{T}_z$ with $z \in \mathbb{R}$.  
For any $h,f \in \mathcal{D}(\mathcal{A}^*)$: 
\begin{footnotesize}
\begin{align}
& \langle \big( \mathcal{A} \mathcal{T}_z^* \mathcal{P} \mathcal{T}_z+ \mathcal{T}_z^* \mathcal{P} \mathcal{T}_z \mathcal{A}^* + \mathcal{B} \mathcal{B} ^*  - \mathcal{T}_z^* \mathcal{P} \mathcal{T}_z \mathcal{C}^* (\mathcal{G} \mathcal{G}^*)^{-1} \mathcal{C}\mathcal{T}_z^* \mathcal{P} \mathcal{T}_z\big) h,  f \rangle \stackrel{(a)}{=} \nonumber \\
& \langle \big( \mathcal{T}_z^* \mathcal{A}  \mathcal{P} \mathcal{T}_z+ \mathcal{T}_z^* \mathcal{P} \mathcal{A}^* \mathcal{T}_z  + \mathcal{T}_z ^* \mathcal{B} \mathcal{B} ^* \mathcal{T}_z  - \mathcal{T}_z^* \mathcal{P}  \mathcal{C}^* (\mathcal{G} \mathcal{G}^*)^{-1} \mathcal{C} \mathcal{P} \mathcal{T}_z\big) h,  f \rangle \hspace{0.05cm}= \nonumber \\
& \langle  \mathcal{T}_z^*  \big(\mathcal{A}  \mathcal{P} +  \mathcal{P} \mathcal{A}^*   +\mathcal{B} \mathcal{B} ^* -  \mathcal{P}  \mathcal{C}^* (\mathcal{G} \mathcal{G}^*)^{-1} \mathcal{C} \mathcal{P} \big) \mathcal{T}_z h,  f \rangle \hspace{2.17cm} = \nonumber \\
& \langle   \big(\mathcal{A}  \mathcal{P} +  \mathcal{P} \mathcal{A}^*   + \mathcal{B} \mathcal{B} ^*  -  \mathcal{P}  \mathcal{C}^*(\mathcal{G} \mathcal{G}^*)^{-1} \mathcal{C} \mathcal{P} \big) \mathcal{T}_z h,  \mathcal{T}_z f \rangle \hspace{2.25cm} = \nonumber \\
& \langle   \big(\mathcal{A}  \mathcal{P} +  \mathcal{P} \mathcal{A}^*   + \mathcal{B} \mathcal{B} ^* -  \mathcal{P}  \mathcal{C}^* (\mathcal{G} \mathcal{G}^*)^{-1} \mathcal{C} \mathcal{P} \big)  h,   f \rangle \hspace{2.85cm} \stackrel{(b)}{=} \nonumber \\
& 0, \nonumber
\end{align}
\end{footnotesize}
where $\stackrel{(a)}{=}$ follows  from Assumptions \ref{ass:A2}, \ref{ass:A3}, and \ref{ass:operators_BCG}.1
 as the operators $\mathcal{A},  \mathcal{B},  \mathcal{C}$ and $\mathcal{G}$ in the ARE are spatially invariant and hence,  they commute with $\mathcal{T}_z$; and $\stackrel{(b)}{=}$ holds because $\mathcal{P}$ satisfies the ARE \eqref{eq:estimation_OARE} by definition. 
Since the ARE \eqref{eq:estimation_OARE} has a \textit{unique} self-adjoint stabilizing solution $\mathcal{P}$ and $\mathcal{T}_z^* \mathcal{P} \mathcal{T}_z$ also satisfies \eqref{eq:estimation_OARE} being self-adjoint and stabilizing,  then $\mathcal {P} = \mathcal{T}_z^* \mathcal{P} \mathcal{T}_z$ and thus,  $\mathcal{P}$ is a spatially invariant operator. 
\end{proof}

\begin{thm}[Spatial invariance of the Kalman filter] 
\label{thm:KB_freq}
Let Assumptions \ref{ass:A1} to \ref{ass:5}
 hold.  Then,  the infinite-dimensional Kalman filter \eqref{eq:KBF_abstract_form} for the spatially invariant plant \eqref{eq:plant_abstract} is equivalently obtained in the spatial frequency domain by solving the following family of finite dimensional AREs parametrized by the spatial frequency $\lambda \in \mathbb{R}$,
 \begin{equation}
    2 \Re(\hat{\mathcal{A}}_{\lambda}) \hat{\mathcal{P}}_{\lambda}  + |\hat{\mathcal{B}}_{\lambda}|^2  - |\hat{\mathcal{C}}_{\lambda}|^2 (|\hat{\mathcal{G}}_{\lambda} |^2)  ^{-1} \hat{\mathcal{P}}_{\lambda}^2 \equiv 0.
    \label{eq:ARE_frequency_almost_everywhere_theorem}
 \end{equation}
 We refer to \eqref{eq:ARE_frequency_almost_everywhere_theorem} as the \textit{power spectral density equation}.
The corresponding Kalman gain operator $\mathcal{L}$ is a spatial convolution and hence,  the filter \eqref{eq:KBF_abstract_form}  is spatially invariant.  
\end{thm}

\begin{proof}
By Assumptions \ref{ass:A2}, \ref{ass:A3}, and \ref{ass:operators_BCG}.1  together with Lemma \ref{lem:spatial_invariance_P},  all the operators in \eqref{eq:estimation_OARE} are spatially invariant. Assumption \ref{ass:5} guarantees the well-posedness of the filtering problem. Taking the spatial Fourier transform of \eqref{eq:estimation_OARE} and using the fact that it diagonalizes spatially invariant operators, together with the  Parseval-Plancherel identity give 
  \begin{align}
      &  \langle \big( \hat{\mathcal{A}}\hat{\mathcal{P}} + \hat{\mathcal{P}} \widehat{\mathcal{A}^*} + \hat{\mathcal{B}}\widehat{\mathcal{B}^*} -  \hat{\mathcal{P}}_{\lambda} \widehat{\mathcal{C}^*} \widehat{(\mathcal{G} \mathcal{G}^*)^{-1}}\hat{\mathcal{C}}\hat{\mathcal{P}} \big) \hat{h},  \hat{f} \rangle = 0,  
\label{eq:estimation_OARE_frequency} 
  \end{align}
$\forall \hat{h}, \hat{f} \in \mathcal{D}(\widehat{\mathcal{A}^*})$. Under density of $\mathcal{D}(\widehat{\mathcal{A}^*})$ in $L^2(\mathbb{R})$,  
\begin{align}
& \big(\hat{\mathcal{A}}\hat{\mathcal{P}} + \hat{\mathcal{P}} \hat{\mathcal{A}}^* + \hat{\mathcal{B}} \hat{\mathcal{B}}^*  - \hat{\mathcal{P}} \hat{\mathcal{C}}^*(\hat{\mathcal{G}} \hat{\mathcal{G}}^*)^{-1} \hat{\mathcal{C}} \hat{\mathcal{P}} \big) \hat{h}   \in \mathcal{D}(\hat{\mathcal{A}^*})^{\bot} = \{0\}.
\label{eq:estimation_OARE_frequency_1}
\end{align}  
Furthermore,  since \eqref{eq:estimation_OARE_frequency_1} holds for any $\hat{h} \in \mathcal{D}(\hat{\mathcal{A}}^*)$,
\begin{align}
& \hat{\mathcal{A}}_{\lambda} \hat{\mathcal{P}}_{\lambda} + \hat{\mathcal{P}}_{\lambda} \hat{\mathcal{A}}_{\lambda}^* + \hat{\mathcal{B}}_{\lambda} \hat{\mathcal{B}}_{\lambda}^* - \hat{\mathcal{P}}_{\lambda} \hat{\mathcal{C}}_{\lambda}^*(\hat{\mathcal{G}}_{\lambda} \hat{\mathcal{G}}_{\lambda}^*)^{-1} \hat{\mathcal{C}}_{\lambda} \hat{\mathcal{P}}_{\lambda} \equiv 0, 
 \label{eq:estimation_OARE_frequency_2}
\end{align}
with $\lambda \in \mathbb{R}. $ For a fixed $\lambda\in \mathbb{R}$, \eqref{eq:estimation_OARE_frequency_2}  is a finite-dimensional ARE with complex coefficients.
Since by Assumption \ref{ass:A1} we work with scalar spatiotemporal fields,  and  $\hat{\mathcal{P}}_{\lambda},  (\hat{\mathcal{G}}_{\lambda} \hat{\mathcal{G}}_{\lambda}^*) = |\hat{\mathcal{G}}_{\lambda}|^2$ and $(\hat{\mathcal{B}}_{\lambda} \hat{\mathcal{B}}_{\lambda}^* ) = |\hat{\mathcal{B}}_{\lambda}|^2$ are real,  \eqref{eq:estimation_OARE_frequency_2} simplifies to \eqref{eq:ARE_frequency_almost_everywhere_theorem}. 
 The Fourier symbol of the Kalman gain $\mathcal{L}$ is 
 \begin{equation}
 \hat{\mathcal{L}}_{\lambda} = \hat{\mathcal{P}}_{\lambda} \hat{\mathcal{C}}^*_{\lambda} (|\hat{\mathcal{G}}_{\lambda}|^2)^{-1},
 \label{eq:Fourier_symbol_KalmanGain}
  \end{equation}
a multiplication operator in the frequency domain.  By the convolution theorem, $\mathcal{L}$ is a spatial convolution.
Thus, the Kalman filter \eqref{eq:KBF_abstract_form} is spatially invariant.
\end{proof}


By Theorem \ref{thm:KB_freq} the operators $\mathcal{P}$ and $\mathcal{L}$ are spatially invariant. Solving the $\lambda$-parameterized 
ARE \eqref{eq:ARE_frequency_almost_everywhere_theorem} explicitly for $\hat{\mathcal{P}}_{\lambda}$ and selecting the solution corresponding to a positive operator yields \eqref{eq:P_L_symbols}.
The spatial invariance of the filter
implies that at each location $x$ the same filter architecture is implemented.
The 
Kalman filter for SIS can be thought of as a $\lambda$-parameterized family of finite-dimensional \textit{modal} Kalman filters: 
$
\frac{\mathrm{d} \hat{\tilde{\psi}}_{\lambda}}{\mathrm{d}t} (t) = (\hat{\mathcal{A}}_{\lambda} - \hat{\mathcal{L}}_{\lambda} \hat{\mathcal{C}}_{\lambda}) \hat{\tilde{\psi}}_{\lambda}(t) + \hat{\mathcal{L}}_{\lambda}\hat{y}_{\lambda}(t), \, \lambda \in \mathbb{R}.
\label{eq:modal_KBF_dynamics}
$

\begin{rem}[Spatial symmetry of $L$] When  Assumption \ref{ass:A4} holds  the Fourier symbol \eqref{eq:Fourier_symbol_KalmanGain_bis} is real.  Since $L$ is a real convolution kernel,
then $L$ must be even in the spatial coordinate. This implies that for a spatial location $x$,  measurements to the left and  right of $x$ are equally valuable for the filtering task at $x$. 
\end{rem}


\subsection{Spatial localization of the Kalman filter for SIS}
\label{subsec:spatial_localization_KF_appendix}

In this subsection, we establish the rapid spatial decay of $\mathcal{L}$.  
 First, we present a useful result in Lemma \ref{lem:analytic_extension}. 


\begin{lem}[Analytic extension of $\hat{\mathcal{L}}_{\lambda}$]
\label{lem:analytic_extension}
 Let Assumptions \ref{ass:A1} to 
 \ref{ass:A4}
 hold and let 
\begin{equation}
\theta \coloneqq \min_i |\Re(z_i)|>0,
\label{eq:theta_def}
\end{equation}
where $z_i$ denote the  branch points  of the 
 extension $\hat{\mathcal{L}}_z$ of the Fourier symbol $\hat{\mathcal{L}}_{\lambda}$ \eqref{eq:Fourier_symbol_KalmanGain_bis} to the complex plane.  
 Then, \\
\indent  i) $\hat{\mathcal{L}}_z$ is analytic in the strip $\mathscr{S}:= \Gamma + \mathbf{i} \mathbb{R}$ of the complex plane, with $\Gamma := (-\theta, \theta)$; 
and \\
\indent ii) 
for any compact $\Gamma_0 \subset \Gamma$, the extension $\hat{\mathcal{L}}_z$ satisfies $|\hat{\mathcal{L}}_z| \leq C (1 + |z|)^N$ for any $z$ such that $\Re(z) \in \Gamma_0$ and appropriate $C,N>0$.
\end{lem}

\begin{proof} 
Under Assumptions \ref{ass:A1} to \ref{ass:5}, the explicit expression \eqref{eq:Fourier_symbol_KalmanGain_bis} for $\hat{\mathcal{L}}_{\lambda}$ holds. We further particularize it to account for Assumption \ref{ass:A4}.


\textit{Part i) Existence of an analytic extension $\hat{\mathcal{L}}_z$ to a strip along the imaginary axis.} Define the extension of \eqref{eq:Fourier_symbol_KalmanGain_bis} to a region of the complex plane by
\begin{small}
\begin{equation}
\hat{\mathcal{L}}_z \coloneqq \big[\Re(\hat{\mathcal{A}}_{\lambda})\big]_{(-\mathbf{i}z)} + \sqrt{\big[\Re(\hat{\mathcal{A}}_{\lambda})^2\  +  \big| \hat{\mathcal{B}}_{\lambda} /\hat{\mathcal{G}}_{\lambda}\big|^2\big]_{(-\mathbf{i}z)}},
\label{eq:analytic_continuation_wannabe}
\end{equation}
\end{small}
where $[\, \boldsymbol{\cdot}\, ]_{(-\mathbf{i}z)}$ denotes substitution of $\lambda \in \mathbb{R}$ by $(-\mathbf{i}z)$,  with $z \in \mathbb{C}$.  
By Assumption \ref{ass:A3},  $\hat{\mathcal{A}}_{\lambda}$ is a polynomial.  Hence,  $[\Re(\hat{\mathcal{A}}_{\lambda})]_{ (-\mathbf{i}z)}$ is an entire function.  Then,  the branch points of \eqref{eq:analytic_continuation_wannabe} are those of the complex square root,  i.e.,  the values of $z$ such that
\begin{subequations}
\begin{align}
\Big[ \Re(\hat{\mathcal{A}}_{\lambda})^2 + \big| \hat{\mathcal{B}}_{\lambda}/\hat{\mathcal{G}}_{\lambda}\big|^2 \Big]_{ (-\mathbf{i}z)} & = 0, \; \text{ or }
\label{eq:BPs_zero}\\
\big| \big[\Re(\hat{\mathcal{A}}_{\lambda})^2 + \big| \hat{\mathcal{B}}_{\lambda}/\hat{\mathcal{G}}_{\lambda}\big|^2 \big]_{(-\mathbf{i}z)} \big| & \to \infty.
\label{eq:BPs_infty}
\end{align}
\label{eq:BPs_proof}
\end{subequations}
\eqref{eq:BPs_proof} determine the analyticity region $\mathscr{S}$ of \eqref{eq:analytic_continuation_wannabe}.  We discard the existence of finite branch points on the imaginary axis.  
By construction,  at $z = \lambda \mathbf{i}$ with $\lambda \in \mathbb{R}$,  the radicand in \eqref{eq:analytic_continuation_wannabe} is $\Re(\hat{\mathcal{A}}_{\lambda})^2 + \big| \hat{\mathcal{B}}_{\lambda}/\hat{\mathcal{G}}_{\lambda} \big|^2$. Following Assumption \ref{ass:operators_BCG}.2 denote $\hat{\mathcal{B}}_{\lambda} = b/p_{\mathcal{B}}(\lambda)$ and $\hat{\mathcal{G}}_{\lambda} = g/p_{\mathcal{G}}(\lambda)$, with $b, g $ nonzero constants and $p_{\mathcal{B}}, p_{\mathcal{G}}$\ polynomials. Then, for \eqref{eq:BPs_zero} to hold for purely imaginary values of $z$, $\Re(\hat{\mathcal{A}}_{\lambda})^2  |g \,p_{\mathcal{B}}(\lambda)|^2 +  |b \, p_{\mathcal{G}}(\lambda)|^2 = 0$. Since both terms in the sum are non-negative, the equality can only be satisfied if both are simultaneously zero. However, that is not possible as by Assumption \ref{ass:operators_BCG}.2 $p_{\mathcal{G}}$ has no real roots.
Proceeding analogously,  by Assumption \ref{ass:operators_BCG}.2  $p_{\mathcal{B}}$ has no real roots, which rules out the existence of purely imaginary finite branch points satisfying \eqref{eq:BPs_infty}.
Thus,  
the finite branch points $z_i$ of $\hat{\mathcal{L}}_z$ have $\Re(z_i) \neq 0$, and
$\hat{\mathcal{L}}_z$ is analytic in the strip
$
\mathscr{S}\coloneqq \big\{ z \in \mathbb{C} \, : \,  |\Re(z)| < \theta \big\},
\label{eq:analyticity_strip}
$
with $\theta$ as defined in \eqref{eq:theta_def}.

\textit{Part ii) $|\hat{\mathcal{L}}_z|$ is polynomially bounded $\forall z$ s.t. $\Re(z) \in \Gamma_0$.} 
\begin{small}
\begin{align}
    | \hat{\mathcal{L}}_z | & := \Big| \big[\Re[\hat{\mathcal{A}}_{\lambda}] \big]_{(-\mathbf{i}z)} + \sqrt{\big[ \Re(\hat{\mathcal{A}}_{\lambda})^2 + \big| \hat{\mathcal{B}}_{\lambda}/\hat{\mathcal{G}}_{\lambda} \big|^2\big]_{(-\mathbf{i}z)}}\; \Big| \nonumber \\
    & \leq
    \underbrace{2 \big| \big[\Re[\hat{\mathcal{A}}_{\lambda}] \big]_{(-\mathbf{i}z)} \big| \vphantom{\big[\hat{\mathcal{W}}_{\lambda}/\hat{\mathcal{V}}_{\lambda} \big]_{\lambda \mapsto (-\mathbf{i}z)} \big|^{\frac{1}{2}}}}_{(a)} + \underbrace{\big| \big[ \, \big| \hat{\mathcal{B}}_{\lambda}/\hat{\mathcal{G}}_{\lambda} \big|^2 \big]_{(-\mathbf{i}z)} \big|^{\frac{1}{2}}  }_{(b)}.
\label{eq:to_bound}
\end{align}
\end{small}
Note that any complex polynomial $p(z) = p_0 + p_1 z  + \dots + p_n z^n $ with $p_i, z \in \mathbb{C}$, satisfies $|p(z)| \leq C (1+|z|)^N$ where
\begin{equation}
    C = \max_{j=0,\dots,n} |p_j| \text{ and } 
    N  = n. 
    \label{eq:coeffs_def}
\end{equation}
$\bullet$ \textit{Bound for $(a)$:}  by Assumption \ref{ass:A3},  
$[\Re(\hat{\mathcal{A}}_{\lambda})]_{(-\mathbf{i}z)} $ is 
a polynomial. Then,
    $
    2 |[  \Re(\hat{\mathcal{A}}_{\lambda})] _{(-\mathbf{i}z)} | \leq C_A (1+|z|)^{N_A},
    \label{eq:i_bound}
    $
    with $C_A$ and $N_A$ chosen following \eqref{eq:coeffs_def}.

\noindent  $\bullet$ \textit{Bound for $(b)$:}  
$
    (b) = \big| \big[ \, \big|\frac{\hat{\mathcal{B}}_{\lambda}}{\hat{\mathcal{G}}_{\lambda}} \big|^2 \big]_{(-\mathbf{i}z)} \big|^{\frac{1}{2}}  = \frac{|b|}{|g|} \ \frac{\big| | p_{\mathcal{G}}(\lambda)|^2_{(-\mathbf{i}z)} \big|^{\frac{1}{2}}}{\big| | p_{\mathcal{B}}(\lambda)|^2_{(-\mathbf{i}z)} \big|^{\frac{1}{2}}}.
$
  Since $\hat{\mathcal{L}}_{z}$ is analytic in $\Gamma$, $\forall z$ s.t. $\Re(z) \in \Gamma_0 \subset \Gamma$ with $\Gamma_0$ compact, $\big| |p_{\mathcal{B}}(\lambda)|^2_{(-\mathbf{i}z)}\big| > \varepsilon$  for some $\varepsilon>0$.  
 Then,
    \begin{small}
    $
        \frac{|b|}{|g|} \ \frac{\big| | p_{\mathcal{G}}(\lambda)|^2_{(-\mathbf{i}z)} \big|^{1/2}}{\big| | p_{\mathcal{B}}(\lambda)|^2_{(-\mathbf{i}z)} \big|^{1/2}}  < 
         \frac{|b|}{|g|} \ \frac{\big| | p_{\mathcal{G}}(\lambda)|^2_{(-\mathbf{i}z)} \big|^{1/2}}{\varepsilon^{1/2}}
        \leq \frac{|b|}{|g|} \big(\frac{p_{max}}{\varepsilon}\big)^{1/2} (1 + |z|)^{\frac{N_p}{2}} 
        \stackrel{(a)}{\leq} \underbrace{\frac{|b|}{|g|} \big(\frac{p_{max}}{\varepsilon}\big)^{\frac{1}{2}}}_{=:C_P} (1 + |z|)^{N_p},
    $
    \end{small}
    where $N_p$ and $p_{max}$ are chosen for the complex polynomial  $|p_{\mathcal{G}}(\lambda)|^2_{(-\mathbf{i}z)}$
     using \eqref{eq:coeffs_def} and $\stackrel{(a)}{\leq} $ holds since $\sqrt{a} \leq a,  \text{ when } a \geq 1$.

Finally,  substitution of the bounds for $(a)$ and $(b)$ in \eqref{eq:to_bound} yields:
$
    \big|\hat{\mathcal{L}}_z \big|  \leq   C_A(1+|z|)^{N_A} + C_P (1 + |z|)^{N_p} \leq C (1+|z|)^N,
    \label{eq:desired_bound}
$
with $C := \max \{ C_A, C_P \}$ and $N:= \max \{ N_A, N_p\}$.
\end{proof} 

\begin{thm}[Spatial decay of $\mathcal{L}$]
\label{thm:decay_rate_L}
Let Assumptions \ref{ass:A1} to \ref{ass:A4}
 hold and let $\theta$ be as defined in \eqref{eq:theta_def}. Then,  
 the Kalman gain  
 $\mathcal{L}$ is such that $e^{-\eta x} \mathcal{L}$  is a tempered distribution,  for every $|\eta|<\theta$. 
\end{thm}
\begin{proof}
Under Assumptions \ref{ass:A1} to \ref{ass:5}, the Fourier symbol $\hat{\mathcal{L}}_{\lambda}$ of $\mathcal{L}$ is \eqref{eq:Fourier_symbol_KalmanGain_bis}. Under Assumption \ref{ass:A4}, its analytic extension $\hat{\mathcal{L}}_z$ satisfies Lemma \ref{lem:analytic_extension}. Then, 
 the Paley-Wiener Theorem \ref{thm:Paley-Wiener} holds and the proof follows by its straightforward application to $\hat{\mathcal{L}}_{\lambda}$.
\end{proof}

\section*{Appendix II: Computation of BP's and analyticity region of \eqref{eq:Swift_Hohenberg_whiteNoise_extension} }
\label{sec:branch_points_Swift_white}
The finite branch points of \eqref{eq:Swift_Hohenberg_whiteNoise_extension} are
\begin{equation}
\zeta_{1,2,3,4,  5,6,7,8} = \pm \sqrt{ \pm 1 - \frac{1}{\Pi_*^2} \pm \mathbf{i}}.
\end{equation}
Due to the symmetries present in the problem,  four of the branch points ($\zeta_{1,2,3,4}$) are located on an inner circunference of radius  $R_{\Omega} = \big((1-\frac{1}{\Pi_*^2} )^2 + 1 \big)^{\frac{1}{4}}$ and the remaining four ($\zeta_{5,6,7,8}$) are located on an outer circunference of radius $R_{\Theta} =\big((1+\frac{1}{\Pi_*^2})^2 + 1 \big)^{\frac{1}{4}}$ in the complex plane.  We explicitly compute the real and imaginary parts of these two sets of branch points.
\begin{itemize}
\item Outer branch points $\zeta_{5,6,7,8}$:
\end{itemize}
Denote $\zeta_5 =  \Theta_r  + \Theta_i \,  \mathbf{i}$ with $\Theta_r, \Theta_i \in \mathbb{R}$.  Thus,  $\zeta_5^2 = \Theta_r^2 - \Theta_i^2 + 2 \Theta_r \Theta_i \, \mathbf{i} = -1 -\frac{1}{\Pi_*^2} + \mathbf{i}$,  from where $\Theta_i = 1/(2 \Theta_r)$.  Denote $\Theta = \Theta_r$.  Then,  $\Theta$ satisfies the biquadratic equation
\begin{equation}
\Theta^4 + \bigg( 1 + \frac{1}{\Pi_*^2}\bigg) \Theta^2 - \frac{1}{4} = 0,
\label{eq:quartic}
\end{equation}
from where $\Theta$ as given in \eqref{eq:Swift_Hohenberg_Theta} is obtained by keeping the inner positive sign when solving \eqref{eq:quartic},  as by definition $\Theta \in \mathbb{R}$.  The corresponding branch points are $\zeta_{5,6,7,8} = \pm \Theta \pm 1/(2 \Theta) \,  \mathbf{i}$. 
\begin{itemize}
\item Inner branch points $\zeta_{1,2,3,4}$:
\end{itemize}
Denote $\zeta_1 = \Omega_r + \Omega_i \, \mathbf{i}$ with $\Omega_r,  \Omega_i \in \mathbb{R}$. 
Proceeding analogously to the previous computation of the outer branch points,  we have that $\Omega_i = 1/(2 \Omega_r)$ and denote $\Omega = \Omega_r$.  The equation for $\Omega$ is then $\Omega^4 + \big( \frac{1}{\Pi_*^2} - 1\big) \Omega^2 - \frac{1}{4} = 0$ which yields $\Omega$ as given in \eqref{eq:Swift_Hohenberg_Omega} and inner branch points $\zeta_{1,2,3,4} = \pm \Omega \, \pm 1/(2 \Omega) \, \mathbf{i}$.  

\begin{figure}[h]
\centering
\includegraphics[width=0.45\linewidth]{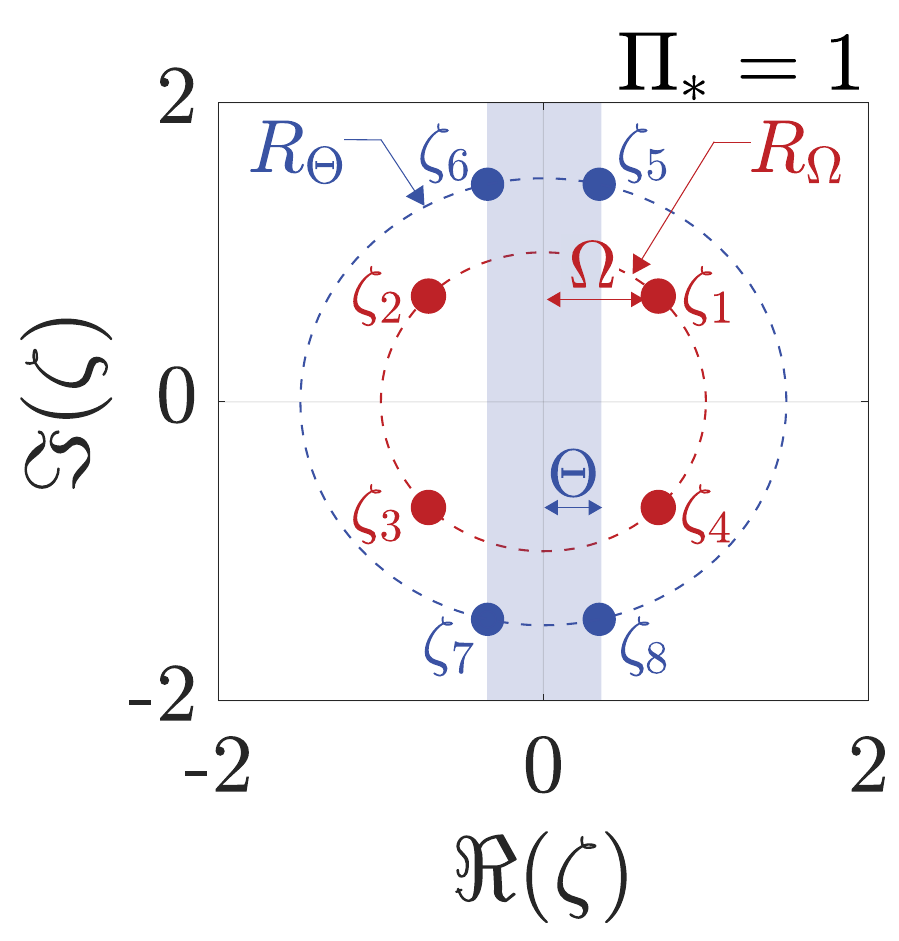}
\caption{\small{Finite inner and outer branch points and analyticity region of the extension \eqref{eq:Swift_Hohenberg_whiteNoise_extension} in the dimensionless complex $\zeta$-plane for $\Pi_* = 1$, where $\Pi_*$ is as defined in \eqref{eq:pi_swift_hohenberg}.  Branch cuts are not represented. }}
\label{fig:Swifthohenber_inner_outer_BPs}
\end{figure}

Using the explicit expressions \eqref{eq:SH_Omega_Theta}, 
it is checked that $\Theta < \Omega, \forall \Pi_*> 0$ (see Fig.  \ref{fig:Swift_whiteNoise}c). Thus, the outer branch points $\zeta_{5,6,7,8}$ define the analyticity strip of \eqref{eq:Swift_Hohenberg_whiteNoise_extension}: although larger in modulus,  their real parts are smaller in absolute value than the respective counterparts of the inner branch points (see Fig. \ref{fig:Swifthohenber_inner_outer_BPs} for an example):  $\Theta$ as given in \eqref{eq:Swift_Hohenberg_Theta} is the dimensionless asymptotic spatial decay rate of the Kalman gain.

\bibliographystyle{IEEEtran}
\bibliography{refs_1}

\begin{IEEEbiography}[{\includegraphics[width=1in,height=1.25in,clip,keepaspectratio]{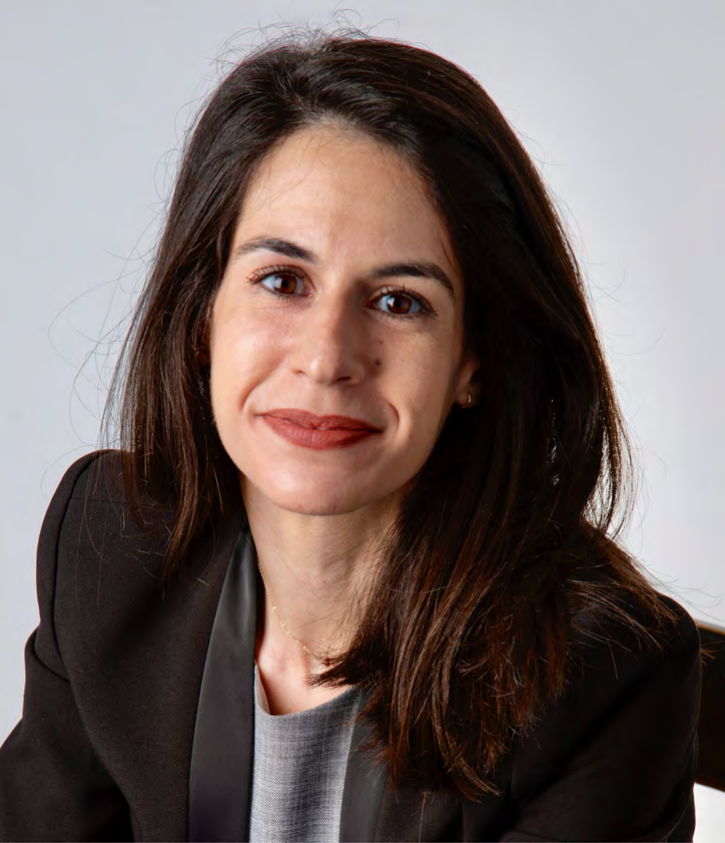}}]{Juncal Arbelaiz}  received B.Sc.  and M.Sc.  degrees in industrial engineering from the University of Navarre, San Sebastian, Guipuzcoa, Spain, in 2014 and 2016, respectively, and the Ph.D. degree in applied mathematics from the Massachusetts Institute of Technology (MIT), Cambridge, MA, USA, in September 2022. She is currently a Postdoctoral Fellow with the Center for Statistics and Machine Learning (CSML).  Her research interests are in optimal decentralized control and estimation of spatially distributed systems. Dr. Arbelaiz is a Schmidt Science Postdoctoral Fellow and was honored as a Rising Star in EECS in 2021. She was the recipient of a Hugh Hampton Young Memorial Fellowship from the Office of Graduate Education at MIT in two consecutive years, 2020 and 2021. She was recognized as a McKinsey Next Generation Women Leader in 2020.  She was also the recipient of a Rafael del Pino Foundation Excellence Fellowship (2019), the Google Women Techmakers Scholarship (2018), the National Award for Academic Excellence of the Government of Spain (2018), the la Caixa Foundation Fellowship (2017) and the Presidential Fellowship from MIT (2016). 
\end{IEEEbiography}
\begin{IEEEbiography}[{\includegraphics[width=1in,height=1.25in,clip,keepaspectratio]{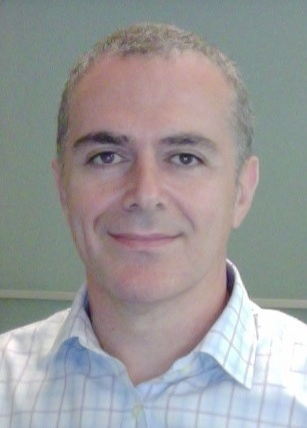}}]{Bassam Bamieh} received the B.Sc. degree in electrical engineering and physics from Valparaiso University, Valparaiso, IN, USA, in 1983, and the M.Sc. and Ph.D. degrees in electrical and computer engineering from Rice University, Houston, TX, USA, in 1986 and 1992, respectively. From 1991 to 1998, he was an Assistant Professor with the Department of Electrical and Computer Engineering, and the Coordinated Science Laboratory, University of Illinois at Urbana-Champaign, Champaign, IL, USA, after which he joined the University of California at Santa Barbara (UCSB), Santa Barbara, CA, USA, where he is currently a Professor of Mechanical Engineering. His research interests include robust and optimal control, distributed and networked control and dynamical systems, shear flow transition and turbulence, and the use of feedback in thermoacoustic energy conversion devices. Dr. Bamieh is a past recipient of the IEEE Control Systems Society G. S. Axelby Outstanding Paper Award (twice), the AACC Hugo Schuck Best Paper Award, and the National Science Foundation CAREER Award. He was elected as a Distinguished Lecturer of the IEEE Control Systems Society in 2005, and a Fellow of the International Federation of Automatic Control (IFAC).
\end{IEEEbiography}
\begin{IEEEbiography}[{\includegraphics[width=1in,height=1.25in,clip,keepaspectratio]{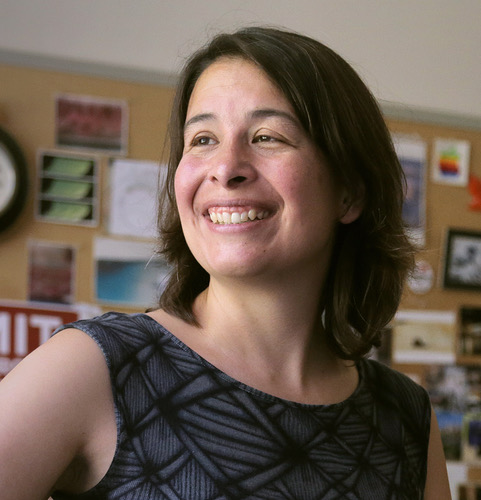}}]{Anette E.  Hosoi} (“Peko”) received an AB in physics from Princeton University, Princeton, NJ, USA in 1992 and an MA and PhD in physics from the University of Chicago,  Chicago,  IL,  USA in 1994 and 1997,  respectively.   She joined the Department of Mechanical Engineering at the Massachusetts Institute of Technology (MIT),  Cambridge,  MA,  USA in 2002 as an assistant professor;  she is currently the Neil and Jane Pappalardo Professor of Mechanical Engineering, Professor of Mathematics, and a core faculty member of the Institute for Data, Systems and Society.   Her research contributions lie at the junction of fluid dynamics, biomechanics, and bio-inspired design.  A common theme in her work is the fundamental study of shape,  kinematic,  and rheological optimization of biological systems with applications to the emergent field of soft robotics.  More recently,  she has turned her attention to problems that lie at the intersection of biomechanics,  applied mathematics,  and sports.  She is the co-founder of the MIT Sports Lab which connects the MIT community with pro-teams and industry partners to address data and engineering challenges that lie within the sports domain.  Dr. Hosoi has received numerous awards including the APS Stanley Corrsin Award,  the Bose Award for Excellence in Teaching,  and the Jacob P. Den Hartog Distinguished Educator Award.  She is a Fellow of the American Physical Society (APS), a Radcliffe Institute Fellow, and a MacVicar Faculty Fellow. 
\end{IEEEbiography}
\begin{IEEEbiography}[{\includegraphics[width=1in,height=1.25in,clip,keepaspectratio]{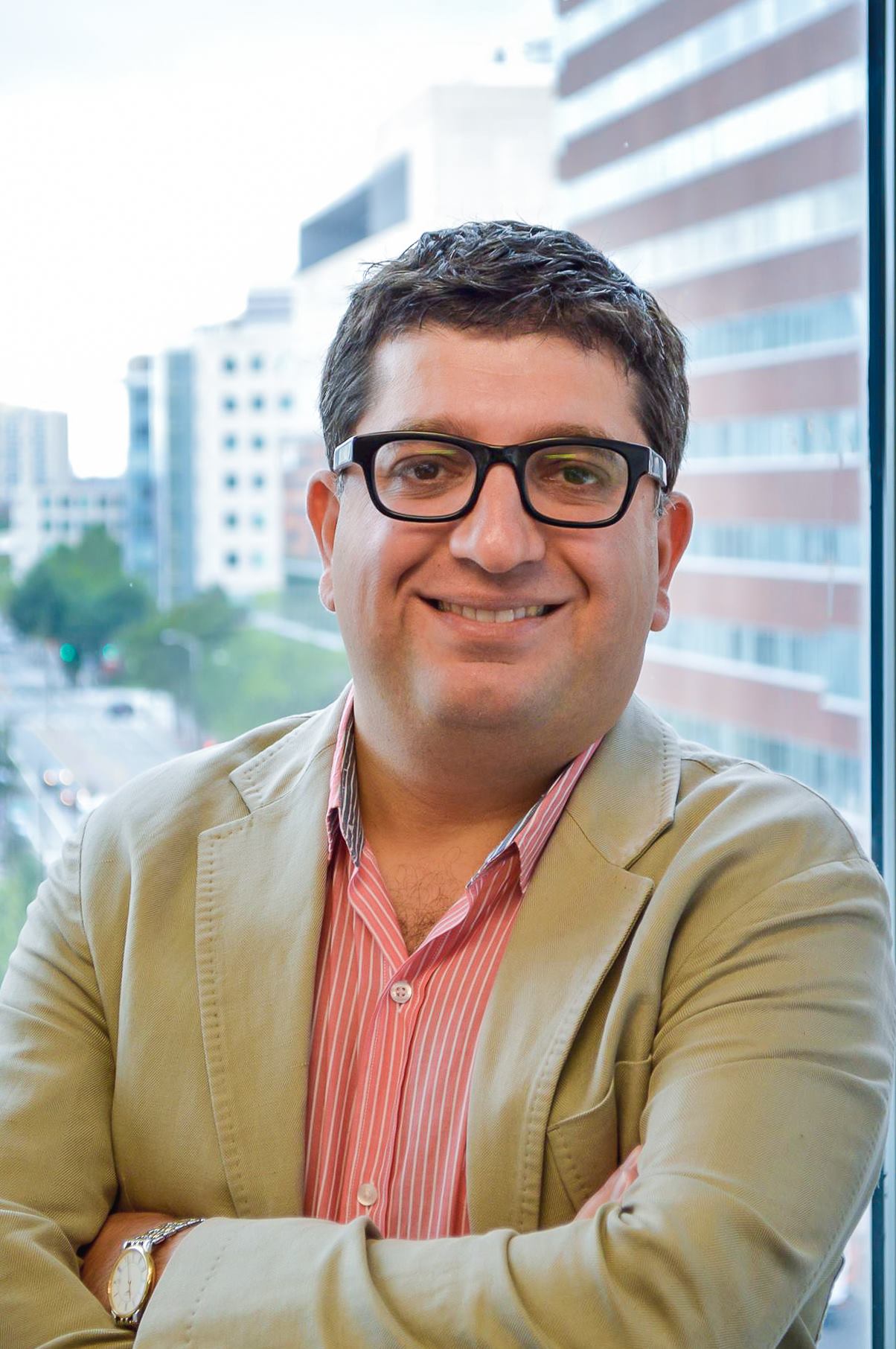}}]{Ali Jadbabaie} is the JR East Professor and Head of the Department of Civil and Environmental Engineering, a core faculty member in the Institute for Data, Systems and Society (IDSS), and a PI at the Laboratory for Information and Decision Systems (LIDS) at the Massachusetts Institute of Technology (MIT). Previously, he served as the director of the Sociotechnical Systems Research Center and as the Associate Director of IDSS at MIT,  which he helped found in 2015.  He received his B.S. from Sharif University of Technology,  his M.S. in electrical and computer engineering from the University of New Mexico,  and his Ph.D.in control and dynamical systems from the California Institute of Technology (Caltech). He was a postdoctoral scholar at Yale University before joining the faculty at the University of Pennsylvania,  where he was subsequently promoted through the ranks and held the Alfred Fitler Moore Professorship in Network Science in the Electrical and Systems Engineering department with secondary appointments in Computer and Information Science and Operations,  Information and Decisions in the Wharton School.  A member of the General Robotics,  Automation, Sensing and Perception (GRASP) Lab at Penn,  Prof.  Jadbabaie was also the cofounder and director of the Raj and Neera Singh Program in Networked and Social Systems Engineering (NETS),  a new undergraduate interdisciplinary degree program.  Prof.  Jadbabaie was the inaugural editor-in-chief of IEEE Transactions on Network Science and Engineering,  a new interdisciplinary journal sponsored by several IEEE societies.  He is a recipient of a National Science Foundation Career Award,  an Office of Naval Research Young Investigator Award, the O. Hugo Schuck Best Paper Award from the American Automatic Control Council,  and the George S.  Axelby Best Paper Award from the IEEE Control Systems Society,  and is a senior author of several student best paper awards.  He is an IEEE fellow and received a Vannevar Bush Fellowship from the office of Secretary of Defense.  His current research interests include the interplay of dynamic systems and networks with specific emphasis on multi-agent coordination and control, distributed optimization, network science, and network economics.
\end{IEEEbiography}
\vfill
\end{document}